\documentclass[traditabstract]{aa}
\usepackage[dutch, english]{babel}
\usepackage{amssymb}
\usepackage{amsmath}
\usepackage{mathrsfs}
\usepackage{latexsym}
\usepackage{longtable}
\usepackage{multirow}
\usepackage{epsf}
\usepackage{color}
\usepackage{graphicx}
\usepackage{float}
\usepackage{enumerate}
\usepackage{txfonts}
\usepackage{natbib}
\usepackage{comment}
\usepackage{capt-of}
\usepackage{rotating}

\newcommand{\one}{~{\sc i}}
\newcommand{\two}{~{\sc ii}}

\newcommand{\mockalph}[1]{}

\title{\object{A resolved, au-scale gas disk around the B[e] star \object{HD 50138}}\thanks{Based on observations performed with X-shooter (program 090.D-0212) and CRIRES (program 084.C-0668), mounted on the ESO \textit{Very Large Telescope}, on Cerro Paranal, Chile, and AMBER mounted on the \textit{Very Large Telescope Interferometer} (programs 082.C-0621, 082.C-0657, 083.C-0144, 084.C-0187, 084.C-0668, 084.C-0983, and 384.D-0482).}}

\authorrunning{L.~E.~Ellerbroek et al.}
\titlerunning{A resolved, au-scale gas disk around the B[e] star \object{HD 50138}}

\author{L.~E.~Ellerbroek \inst{1}
\and
M.~Benisty\inst{2}
\and
S.~Kraus\inst{3}
\and
K.~Perraut\inst{2}
\and
J.~Kluska\inst{2}
\and
J.~B.~Le~Bouquin\inst{2}
\and
M.~Borges Fernandes\inst{4}
\and
A.~Domiciano de Souza\inst{5}
\and
K.~M.~Maaskant\inst{6,1}
\and
L.~Kaper\inst{1}
\and
F.~Tramper\inst{1}
\and
D.~Mourard\inst{5}
\and
I.~Tallon-Bosc\inst{7}
\and
T.~ten~Brummelaar\inst{8}
\and
M.~L.~Sitko\inst{9,10,}\thanks{Visiting Astronomer, Infrared Telescope Facility, operated by the University of Hawaii under Cooperative Agreement no. NNX-08AE38A with the National Aeronautics and Space Administration, Science Mission Directorate, Planetary Astronomy Program.}
\and
D.~K.~Lynch\inst{11,12,\star\star}
\and
R.~W.~Russell\inst{11,\star\star}
}

\institute{Anton Pannekoek Institute, University of Amsterdam, Science Park 904, 1098 XH Amsterdam, The Netherlands\\
\email{lucas.ellerbroek@gmail.com}
\and
Universit\'{e} Grenoble Alpes, IPAG, F-38000 Grenoble, France\\
CNRS, IPAG, F-38000 Grenoble, France
\and
School of Physics, University of Exeter, Stocker Road, Exeter EX4 4QL, UK
\and
Observat\'{o}rio Nacional, Rua General Jos\'{e} Cristino 77, 20921-400 S\~{a}o Cristov\~{a}o, Rio de Janeiro, Brazil
\and
Laboratoire Lagrange, UMR 7293 UNS-CNRS-OCA, Boulevard de lÕObservatoire, CS 34229, 06304 NICE Cedex 4, France
 \and
Leiden Observatory, Leiden University, PO Box 9513, 2300 RA Leiden, The Netherlands
\and
Universit\'{e} de Lyon, 69003 Lyon, France; Universit\'{e} Lyon 1, Observatoire de Lyon, 9 avenue Charles Andr\'{e}, 69230 Saint Genis Laval; CNRS, UMR 5574, Centre de Recherche Astrophysique de Lyon; Ecole Normale Sup\'{e}rieure, 69007 Lyon, France
\and
The CHARA Array of Georgia State University, Mount Wilson Observatory, 91023 Mount Wilson CA, USA
\and
Department of Physics, University of Cincinnati, Cincinnati OH 45221, USA
\and
Space Science Institute, 4750 Walnut Street, Boulder, CO 80303, USA
\and
The Aerospace Corporation, Los Angeles, CA 90009, USA
\and
Thule Scientific, Topanga, CA 90290, USA
}

\date{Received; accepted}

\abstract{HD~50138 is a B[e] star surrounded by a large amount of circumstellar gas and dust. Its spectrum shows characteristics which may indicate either a pre- or a post-main-sequence system. Mapping the kinematics of the gas in the inner few au of the system contributes to a better understanding of its physical nature. We present the first high spatial and spectral resolution interferometric observations of the Br$\gamma$ line of HD~50138, obtained with VLTI/AMBER. The line emission originates from a region more compact (up to $3$~au) than the continuum-emitting region. Blue- and red-shifted emission originates from the two different hemispheres of an elongated structure perpendicular to the polarization angle. The velocity of the emitting medium decreases radially. An overall offset along the NW direction between the line- and continuum-emitting regions is observed. We compare the data with a geometric model of a thin Keplerian disk and a spherical halo on top of a Gaussian continuum. Most of the data are well reproduced by this model, except for the variability, the global offset and the visibility at the systemic velocity. The evolutionary state of the system is discussed; most diagnostics are ambiguous and may point either to a post-main-sequence or a pre-main-sequence nature.}


\begin{document}

\maketitle

\section{Introduction}
B[e] stars are an enigmatic class of stellar objects, the nature of which is in many cases unknown and strongly debated. They are defined as stars with spectral type B that show forbidden emission lines in their optical spectra, as well as a strong near-infrared excess \citep{Slettebak1976, Allen1976, Zickgraf1998}. The forbidden lines originate in a tenuous circumstellar medium, while a dust envelope or a disk radiates in the infrared. The denomination ``B[e] star" is phenomenological; the defining characteristics can be produced by a heterogeneous set of astrophysical objects. Among its members are both young (pre-main-sequence stars) and evolved systems (e.g.,  supergiants, interacting binaries, and planetary nebulae). For many systems, determining their configuration and evolutionary state proves to be a difficult observational challenge \citep{Lamers1998, Miroshnichenko2007}.

\citet{Lamers1998} formulated a classification scheme for B[e] stars of different nature based on their  spectral lines, suspected luminosity and environment. For some objects, however, the diagnostics in Lamers' scheme are inconclusive. These systems may be better understood by combining observations with high resolution in the spectral, spatial and temporal domain. In the last decade, a new generation of optical/near-infrared interferometers equipped with high spectral resolution instruments have become available. With these, circumstellar gas dynamics can be mapped with unprecedented spatial (milli- to micro-arcsecond) and spectral ($\Delta \varv \sim 25$~km~s$^{-1}$) resolution. This has been a successful method to resolve some of the most intensely debated B[e] systems \citep{Malbet2007, Domiciano2007, Millour2009, Weigelt2011, Kraus2008b, Kraus2012, Wang2012, Wheelwright2012a, Wheelwright2012c, Wheelwright2013}. In most of these cases, binary interaction is the most probable cause of the complex circumstellar environment. Gas disks are found to dominate the emission in the inner few astronomical units (au); only in a few cases a Keplerian velocity field could be resolved \citep[e.g.,][]{Kraus2012}. 


In this paper, we present the first spectro-interferometric study employing high spectral resolution of the puzzling B[e] star HD~50138 (V743~Mon, MWC~158). Being among the brightest B[e] stars in the sky, it is located at a distance of $500\pm150$~pc \citep{VanLeeuwen2007}  and has not been associated with a star-forming region. It may be part of the Orion-Monoceros molecular cloud complex  \citep{Maddalena1986}, but because of the uncertainty in the distance, this cannot be confirmed. 

Despite the ample amount of observations and literature, no definitive conclusion has been drawn regarding its evolutionary state. Arguments have been made to classify it as a pre-main-sequence object \citep{Morrison1995} or a star on, or just evolving off the main-sequence \citep[][BF09]{Borges2009}. For more discussion on its evolutionary state, see e.g., \citet{Jaschek1993, Jaschek1998, Lamers1998}. The main property favoring a pre-main-sequence nature are spectral infall signatures. Conversely, the occurrence of shell phases have been interpreted as signs of a post-main-sequence nature. Many characteristics are ambiguous, like the isolation of the object, its large infrared excess and possible binarity \citep{Cidale2001, Baines2006}. 


The circumstellar dust around the system is distributed in an aspherical geometry, well represented by a moderately inclined disk, $i=56\pm4^\circ$, as determined by \citet[][BF11]{Borges2011} based on near- (au-scale) and mid-infrared (10 au-scale) interferometry. The same authors find a position angle (from north through east) of the disk major axis, $\psi=71\pm7^\circ$, perpendicular to the polarization angle \citep[$159\pm4^{\circ}$,][]{Bjorkman1998, Yudin1998, Oudmaijer1999}. Signatures of outflowing and infalling gas are found in emission lines (\citealt{Morrison1995}; \citealt{Grady1996}; \citealt{Pogodin1997}; BF09; \citealt{Borges2012}, BF12). Spectropolarimetry by \citet{Bjorkman1998} suggests that a geometrically thin gas disk exists, where electron scattering dominates the polarization. Other polarimetry studies also find evidence for a circumstellar rotating disk or equatorial outflow \citep{Oudmaijer1999, Harrington2007, Harrington2009}.

Spectro-astrometric measurements of the H$\alpha$ line show a time-dependent shift of the photocenter and a decreasing spatial width across the line  \citep{Baines2006}. This is interpreted by the authors as being a sign of a binary companion on a wide orbit. The circumstellar material may be the result of the interaction with a much closer companion. Given the limited time sampling of monitoring campaigns to date, no evidence for a spectroscopic binary has been found (\citealt{Corporon1999}; BF12). Photometric and spectroscopic variability is detected on timescales from days to years from as early as the 1930s. This has been attributed to shell phases and outbursts, during which the ejected material absorbs and scatters radiation in lines and in the continuum. This variability remains an actively discussed phenomenon \citep[][BF09, BF12]{Merrill1931, Merrill1933, Doazan1965, Hutsemekers1985, Andrillat1991, Halbedel1991, Bopp1993, Pogodin1997}.

In this paper, we focus on the kinematics of the Br$\gamma$ emission line tracing the circumstellar gas around HD~50138. We present a series of observations performed with the \textit{Astronomical Multi-Beam Combiner} (AMBER) on the \textit{Very Large Telescope Interferometer} (VLTI), with the VLT \textit{Cryogenic High-Resolution Infrared Echelle Spectrograph} (CRIRES) and VLT/X-shooter, all located on Cerro Paranal, Chile. We supplement the dataset with observations of the H$\alpha$ line, taken with the \textit{Visible Spectrograph and Polarimeter} (VEGA) on the CHARA array on Mount Wilson, California, USA. The observations are described in Sect.~\ref{sec:observations}. We present the optical-to-infrared stellar and circumstellar spectra, and the most important trends and signatures found in the AMBER, CRIRES and VEGA data in Sect.~\ref{sec:results}. With baselines up to 120~m and a high spectral resolution ($\varv=25$~km~s$^{-1}$), we are able to trace photocenter shifts at a few micro-arcseconds resolution in the line emission. In Sect.~\ref{sec:modeling}, the interferometric observations are compared to a geometric model, which consists of a Keplerian disk and halo. We discuss the validity of this and other possible model geometries in Sect.~\ref{sec:discussion}. The implications for the object's evolutionary state are also considered. The main conclusions of this work are presented in Sect.~\ref{sec:conclusion}.

\begin{figure}[!t]
   \centering
   \includegraphics[width=0.75\columnwidth]{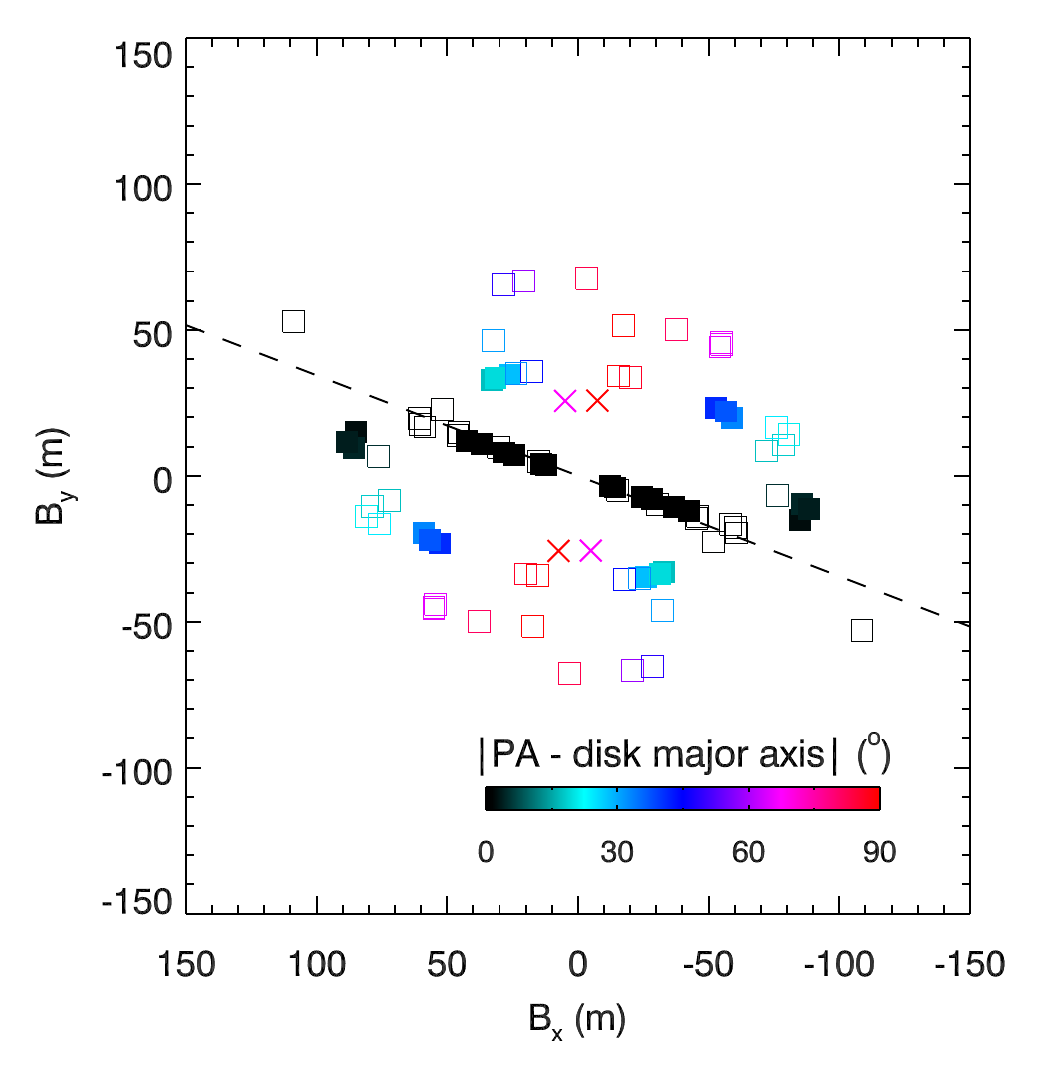}
   \caption{Coverage of the $(u,v)$-plane of the VLTI/AMBER-observations with $\mathcal{R}\sim12,000$ (filled squares) and $\mathcal{R}\sim1500$ (open squares). The two CHARA/VEGA-observations are also displayed (crosses). The dashed line is aligned with the major axis of the modeled disk ellipse ($\psi=71^\circ$). Colors indicate the baseline PA with respect to this line.}
   \label{fig:uvcoverage}
\end{figure}

\begin{table*}[t]
\centering
\caption{Journal of the observations.}
\label{tab:obs}
\scriptsize{
\begin{tabular}{lllllll}
& Date  & Config. & $B_{\rm proj}$ (m) & PA ($^\circ$) & $\mathcal{R}$ (eff.)  & Calibrator \\
\hline
\hline
\multicolumn{6}{c}{\textit{VLTI/AMBER}}\\
\hline
1 & 2009-02-28  & E0G0H0 &   15 / 29 / 44  & 74 / 74 / 74 &12,000 & HD52938\\
2 & 2009-04-25   & E0G0H0 &  13 / 26 / 38 &  74 / 74 / 74 &12,000 & HD52938\\
3 & 2010-01-31   & U2U3U4 &  43 / 62 / 86 & 37 / 108 / 80 & 12,000 & HD44891\\
4 & 2010-03-03   & U2U3U4 & 45 / 58 / 86 & 45 / 114 / 84 & 12,000 & HD44891\\
5 & 2010-03-04   & U2U3U4 & 46 / 61 / 89 & 43 / 111 / 83 & 12,000 & HD44891\\
6 & 2009-01-16    & D0G1H0 & 56 / 68 / 71  & 66 / 177 / 129  & 1500 & HD52938\\
7 & 2009-04-30  & G1D0 &  54  & 161  &1500 & HD52938\\
8 & 2009-11-15   & E0G0H0 &   16 / 32 / 48 & 72 / 72 / 72  &1500 & HD53267\\
9 & 2009-12-04   & U1U2U4 &  57 / 77 / 121 & 35 / 85 / 64  &1500 & HD44891\\
10 & 2010-02-07   & D0G1 &  63 & 143  & 1500 & HD52938\\
11 & 2013-11-12   & D0G1I1 &  70 / 40 / 72 & 129 / 26 / 97 & 1500 & HD59881\\
\hline

\multicolumn{6}{c}{\textit{CHARA/VEGA}}\\
\hline
& 2010-10-11 & S1S2 & 26 & 11 & 160 & HD46487\\
& 2012-10-29 & S1S2 &  27 & --16 & 160 & HD46487\\
\hline
\multicolumn{6}{c}{\textit{VLT/CRIRES}}\\
\hline
& 2009-11-05 & & & 15 / 75 / 135 & 100,000 & HD60803\\
\hline
\multicolumn{6}{c}{\textit{VLT/X-shooter}}\\
\hline
 & 2013-02-14 & & & parallactic ang. & \multicolumn{2}{l}{9100--17,400} \\
\end{tabular}
}
\end{table*}

\section{Observations and data reduction}
\label{sec:observations}

A summary of the observations which are presented in this paper is given in Table~\ref{tab:obs}.

\subsection{VLTI/AMBER Spectro-interferometry}
\label{sec:observations:amber}

We observed HD~50138 between January 2009 and November 2013, during eleven nights.  We  used the near-infrared  instrument AMBER on the VLTI \citep{scholler07, Haguenauer2010}. AMBER enables the simultaneous combination of three beams  in the $K$ band (2.0-2.4~$\mu$m), with a spectral resolving power up to $\mathcal{R} \sim$12,000 \citep{petrov07}. 


Six measurements have been obtained at medium spectral resolution ($\mathcal{R}\sim1500$, $\Delta \varv \sim 200$~km~s$^{-1}$) and five at high resolution ($\mathcal{R}\sim12,000$, $\Delta \varv \sim 25$~km~s$^{-1}$). We performed these  observations using the relocatable 1.8~m auxiliary telescopes (ATs)  and the 8.2~m unit telescopes (UTs), both arrays in two different configurations. The longest projected baseline is $\sim$121~m. Fig.~\ref{fig:uvcoverage} displays the ($u,v$)-plane coverage of the observations. The position angles (PA) of the baselines are color-coded in this image. The broadest range in spatial scales was intentionally achieved along a PA of $\sim 71^\circ$, parallel to the disk major axis as found by BF11, where the object is most extended.

Each measurement of HD~50138 was preceded and followed by observations of calibrator targets to measure the instrumental transfer function and to correct  for  instrumental effects.  We used the following calibrators: HD52938 (angular diameter $\theta\sim 0.87 \pm 0.01$~mas), HD53267 ($\theta\sim 0.86\pm0.01$~mas), HD44891 ($\theta\sim 1.41\pm0.02$~mas), and HD59881 ($\theta\sim 0.44\pm0.03$~mas). The angular diameters were based on uniform disk models. The diameters of the first two calibrators were taken from \citet{Merand2005}; the latter two were directly computed by SearchCal (JSDC2 catalog, \citealt{Bourges2014}). All the observations  were performed using the fringe-tracker FINITO \citep{lebouquin08}. 

The  data  reduction was performed  following  standard procedures described in \citet{Tatulli2007a} and \citet{chelli09},   using  the \texttt{amdlib} package, release  3.0.6, and the \texttt{yorick} interface provided by the Jean-Marie Mariotti Center (JMMC)\footnote{The calibrated data in the OI-FITS format \citep{pauls05} will be included in the JMMC database \texttt{http://www.jmmc.fr}.}. Raw spectral visibilities, differential phases, and closure phases were extracted for all the  frames of each  observing file. The frames of 5 to 15 consecutive observations were merged and a global selection of 80\% of the highest quality frames was made to achieve higher accuracy on the visibilities and phases. The UV coverage didn't change significantly during these sequences. The transfer function was obtained by averaging the calibrator measurements (from 3 to 5 observing blocks, depending on the night), over the entire sequence of observations, after correcting for their intrinsic diameters (Uniform Disk). The VLTI field of view is $\sim$300 and $\sim$60 mas for the AT and UT observations, respectively. 


The wavelengths were converted to a velocity scale. The systemic velocity $\varv = 0$ corresponds to the center of the Br$\gamma$ line, determined from the spectrum by a single Gaussian fit. From the X-shooter spectrum we determined that the Br$\gamma$ line center is consistent with the systemic velocity as determined from the photospheric lines (BF09). The absolute value of the visibilities obtained with the UT baselines could not be determined because of vibrations of the telescopes. These vibrations randomly affect the observations with time, and do not necessarily alter the calibrator and science target observations in the same way. Unreliable calibration may also come from different FINITO locking performances on the calibrator and the science target. However, this issue affects all spectral channels in the same way, and does not modify our conclusions, most of which are based on differential quantities. 


\subsection{CHARA/VEGA Spectro-interferometry}
\label{sec:observations:vega}

We observed HD~50138 around the H$\alpha$ line using the visible VEGA spectrograph (\cite{Mourard2009} at the CHARA array \citep{TenBrummelaar2005}. We observed in the medium resolution mode ($\mathcal{R}=5000$), with the S1S2 baseline ($B_{\rm p} \sim$~28~m). We operated VEGA in parallel with the CLIMB near-infrared beam combiner acting as a coherence sensor \citep{Sturmann2010}.  We measured a typical residual jitter on the optical path difference of the order of 7~$\mu$m. We followed a sequence calibrator-target-calibrator, with 40 or 60 blocks of 1000 short exposures (of 25 ms) per star, using two calibrators: HD 46487 ($\theta\sim 0.178 \pm 0.013$~mas) and HD 59881 ($\theta\sim 0.42 \pm 0.03$~mas). 

The spectra were extracted at $\mathcal{R}=5000$, using a classical scheme of collapsing the 2D flux in one spectrum, calibrating the pixel-wavelength relation with a Thorium-Argon lamp, and normalizing the continuum by a polynomial fit.  As HD~50138 is close to the limiting magnitude of the VEGA instrument, we had to reduce the spectral binning to $\mathcal{R}=160$ to compute differential visibilities and phases \citep{Mourard2009}. The errors on these quantities correspond to the root-mean-square variability in the continuum. The error values in the pixels containing the line emission were diminished by a factor $\sqrt{2}$ to account for the increased flux, which is around 2 times higher in the line with respect to the continuum.

\subsection{VLT/X-shooter Spectroscopy}
\label{sec:observations:xshooter}

Spectra of HD 50138 were obtained on 2013-02-14, UT 04:53, with X-shooter on the VLT. X-shooter covers the optical to near-infrared spectral region in three separate arms: UVB (290--590~nm), VIS (550--1010~nm) and NIR (1000-2480~nm; \citealt{Vernet2011}). Narrow slits were used: $0\farcs5$, $0\farcs4$ and $0\farcs4$ in the three spectrograph arms, respectively. This resulted in a spectral resolving power $\mathcal{R} \equiv \lambda/\Delta\lambda$ of 9100 in UVB, 17,400 in VIS and 11,300 in NIR. The S/N was 120 at 450~nm and 55 at 2150~nm.

The frames were reduced using the X-shooter pipeline \citep[version 1.5.0,][]{Modigliani2010}, employing the standard steps of data reduction, i.e. order extraction, flat fielding, wavelength calibration and sky subtraction. The wavelength calibration was verified by fitting selected OH lines in the sky spectrum. Flux-calibration was performed using spectra of the spectrophotometric standard star GD0.653 (a DA white dwarf). The slit losses were estimated from measuring the seeing full width at half maximum (FWHM, $\sim0\farcs9$ in $V$) from the spatial profile of the point source on the frame. These estimates were refined by comparing the obtained spectral energy distribution (SED) to the averaged photometry (\citealt{Sitko2004}, see Sect.~\ref{sec:sed}). This procedure introduces an uncertainty of about 10\% in the absolute flux calibration; the relative flux calibration is accurate to within 3\%. 

The wavelengths and velocities used throughout this paper are expressed in the systemic rest frame, for which we adopt $35$~km~s$^{-1}$ with respect to the Local Standard of Rest (BF09).

\subsection{VLT/CRIRES Spectro-astrometry}
\label{sec:observations:crires}

Spectra of HD~50138 were obtained in the $K$-band with VLT/CRIRES, for the purpose of spectro-astrometry of the Br$\gamma$ line. The data were taken on 2009-11-05 at UT~06:15. A slit of $0\farcs2$ was used, resulting in a spectral resolving power $\mathcal{R} = 100,000$. The calibrator HD~60803 was observed directly afterwards. The rotator position angles were at 15$^\circ$, 75$^\circ$ and 135$^\circ$, and the respective counter-parallel angles. For a detailed description of the data reduction procedure, see \citet{Kraus2012}.

\subsection{Supplementary data}

Additional spectroscopic and photometric data from previous studies and data archives are used in this paper. An optical high-resolution ($R\sim80,000$) spectrum was taken in March 2007 with the Narval spectropolarimeter at the telescope Bernard Lyot at the observatory of Pic du Midi, France. This spectrum was also presented in BF09. 

Mid-infrared spectra were obtained on 1999-12-24 and 2003-01-08 with The Aerospace Corporation's \textit{Broad-band Array Spectrograph System} (BASS) at the Infrared Telescope Facility \citep{Sitko2004}. This instrument covers the 3--13~$\mu$m wavelength region. BASS is described more fully in \citet{Sitko2008}. Magnitudes from the IRAS observatory are also used.

\begin{figure}[!t]
   \centering
   \includegraphics[width=0.99\columnwidth]{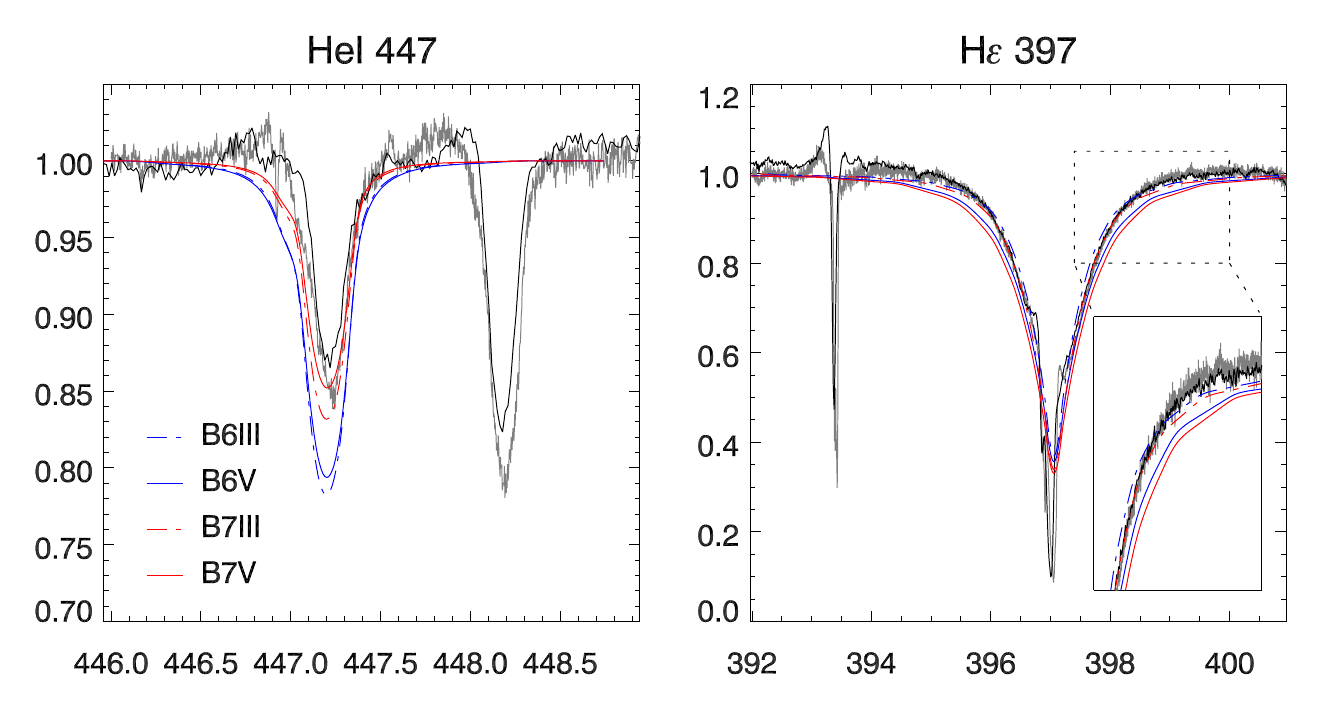}
   \caption{Spectra of the He\one~$\lambda$447.1~nm (\textit{left}) and H$\epsilon$ (\textit{right}) lines, obtained with X-shooter (black line) and Narval (gray line). FASTWIND model profiles are overplotted for B6 and B7 giants and main-sequence stars. The inset in the right panel shows a detail of the line wing. The B7 III model provides the best fit with the observed profiles. The Ca\two~H~$\lambda$393~nm and Mg\two~$\lambda$448~nm lines are also visible.}
   \label{fig:type}
\end{figure}

\section{Results}
\label{sec:results}

\subsection{Spectroscopy}
\label{sec:results:spectroscopy}

In this section, we present spectra obtained with VLT/X-shooter and the AMBER Br$\gamma$ spectra. We compare these to the Narval high-resolution optical spectrum obtained in 2007. 



The spectral type of HD~50138 is difficult to constrain because of the temporal variability. Types in the range B5-A0 and luminosity classes I-V have been proposed in the literature; see references in BF09. The range in spectral types likely is the result of temporal variations (BF09, BF12). BF09  determine a spectral type of B6-B7~III-V based upon a detailed analysis of photometric and spectroscopic data. We check this estimate for consistency by comparing the X-shooter spectra to model spectra of these subtypes. We adopt a similar by-eye fitting method to the one described in \citet{Ochsendorf2011}, which is to qualitatively compare the observed profiles of selected lines to model spectra of B-type stars, varying over a range in temperature and surface gravity.  

For mid- to late-B type spectra, the main temperature diagnostic is provided by the He\one~$\lambda$447~nm line, while the luminosity class is determined from the wings of the H$\epsilon$ line \citep{Gray2009}. Model profiles for these two lines are calculated for the subtypes B6-B7~III-V, using the non-LTE radiative transfer code FASTWIND \citep{Puls2005}. We adopt $\varv \sin i = 100$~km~s$^{-1}$ (BF12). The modeled spectra are compared to the observed profiles in Fig.~\ref{fig:type}; the B7~III model has the best overall fit and we adopt this model in the remainder of the paper. The spectral type is consistent with the distance ($d=500$~pc) and extinction ($A_V=0.4$~mag) of the source (see e.g. BF09). 

\begin{figure}[!t]
   \centering
   \includegraphics[width=0.7\columnwidth]{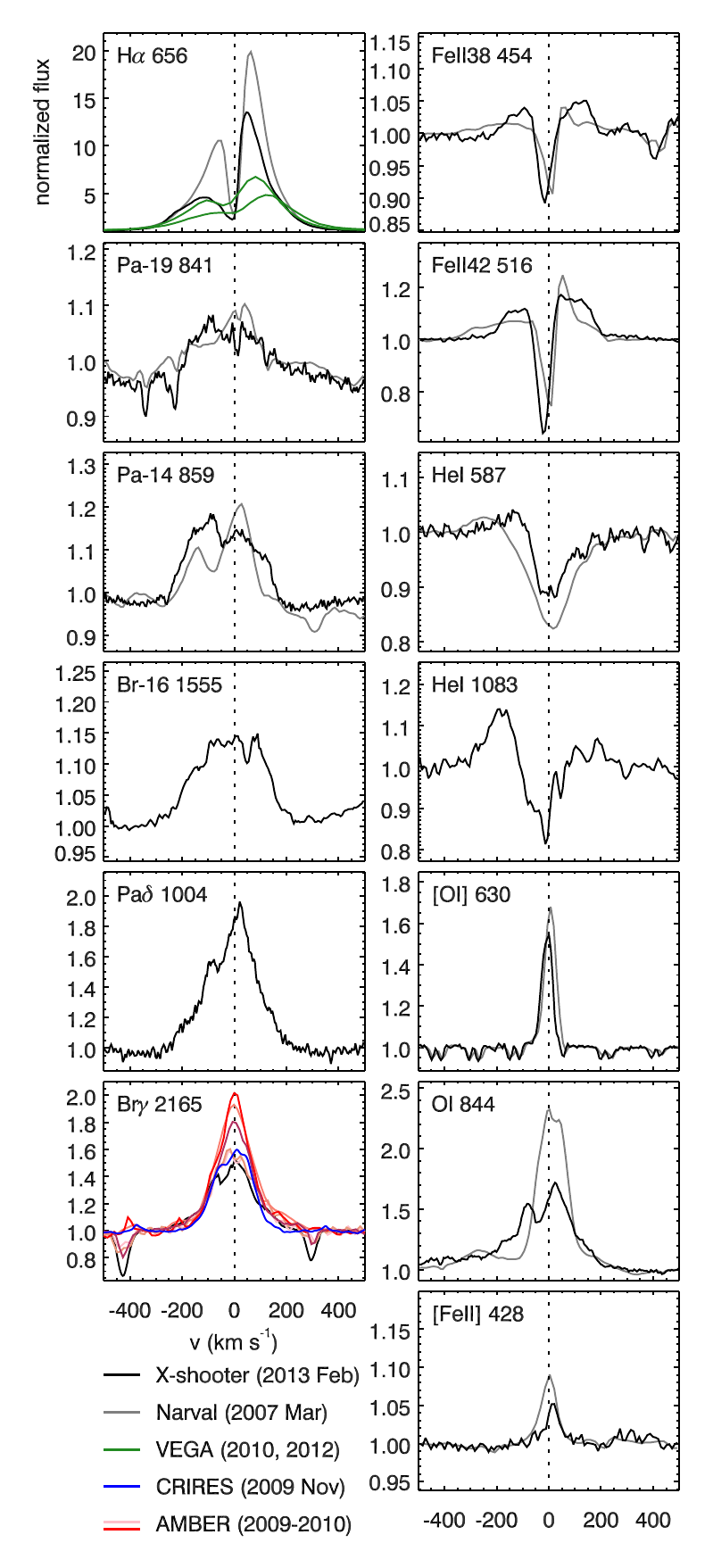}
   \caption{Selection of spectral lines of HD~50138 obtained with X-shooter, Narval (BF09), and AMBER. All lines show some degree of variability. Note the varying strength of double peaks and the component at $\varv\sim0$~km~s$^{-1}$. The Narval and CRIRES spectra are rebinned to match the X-shooter resolution (9100 in the optical, 17,400 in the NIR).}
   \label{fig:xshooter}
\end{figure}




The circumstellar gas environment produces many emission lines; nearly 300 are detected in the X-shooter range. A selection representative of the different line morphologies seen across the X-shooter range is displayed in Fig.~\ref{fig:xshooter}. Additional spectra are also plotted to illustrate variability on timescales from days to years. These data are the AMBER spectra of Br$\gamma$ ($\mathcal{R}=12,000$), the CRIRES spectrum of Br$\gamma$ ($\mathcal{R}=100,000$) and the Narval optical spectrum. The resolution of the high-resulution spectra has been degraded to match the X-shooter spectrum.

All profiles for which multiple observations are available show some degree of variability. The H$\alpha$ profile is double-peaked; the red-shifted peak is strongest. The blue-to-red peak intensity ratio is documented to vary between 0.3 and 0.9 (see BF09 and references therein). The other H~{\sc i} profiles also show double peaks; their blue-to-red variability is not correlated with the H$\alpha$ variability. The higher Paschen and Brackett transitions have relatively shallow profiles with pronounced peaks. The double peaks in the lower transitions, like H$\alpha$ and Pa$\delta$, are less pronounced; they are ``filled in" by an additional broad ($\Delta\varv\sim60$~km~s$^{-1}$) and sharply peaked component at zero velocity. This is also seen in Br$\gamma$, where the profile changes from double- to single-peaked between the different AMBER and X-shooter observations. An abrupt change from single- to double-peaked is also seen in the O\one~$\lambda$844~nm line.

The Fe~{\sc ii} lines have an emission component with a broad FWHM, as well as a narrow, variable absorption component close to zero velocity, which has been attributed to cold material in, e.g., a circumstellar shell or halo (see e.g. \citealt{Pogodin1997}, BF09). In the high-resolution data, it can be seen that the absorption is a blend of multiple narrow components (BF09). The He~{\sc i} lines (as well as optical Mg~{\sc ii} and Si~{\sc ii} lines, that are not shown) have a central absorption blended with blueshifted emission around $-100$~km~s$^{-1}$. This could be an absorption profile combined with emission in a wind or outflow approaching the observer. It has also been interpreted as an inverse P-Cygni profile indicative of infall \citep{Morrison1995, Pogodin1997}. Which of these two interpretations is correct depends on the value of the systemic velocity; with the adopted value from BF09 the first interpretation is favored.  

The SED displayed in Fig.~\ref{fig:sed} is compiled from the X-shooter spectrum and infrared data from \citet{Sitko2004}. A 13,000~K, $\log g=3.5$ atmospheric model \citep[consistent with a B7~III spectral type][]{Kurucz1993} is overplotted, reddened with an $A_V=0.4, R_V=3.1$ extinction law \citep{Cardelli1989}. The resulting stellar radius and luminosity are $R_*=7.0\pm2.1$~R$_\odot$ and $L_*=(1.2\pm0.4)\times10^3$~L$_\odot$. The stellar-to-total flux ratio $f=F_*/F_{\rm tot}$ in the $K$-band provides an important constraint for the analysis of the interferometric data (see Sect.~\ref{sec:results:size} and ~\ref{sec:results:spectroastrometry}). From the SED fit, we obtain $f=0.08\pm0.01$. 

\begin{figure*}[!h]
   \centering
   \includegraphics[height=0.85\textwidth, trim=0cm 13.5cm 0cm 0cm, clip=true, angle=270, origin=c]{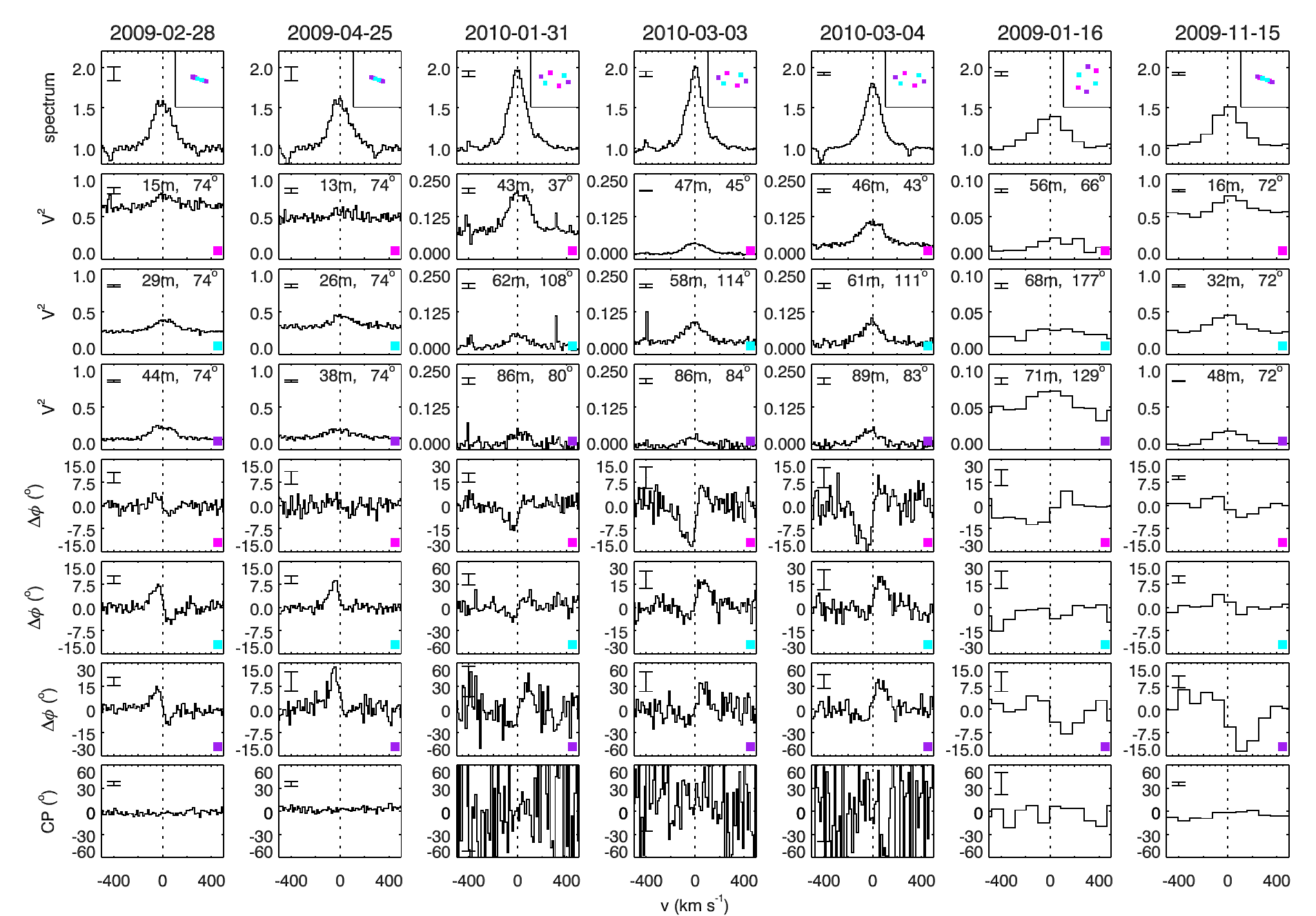}
   \caption{Spectra, squared visibilities, differential phases and closure phase of the Br$\gamma$ line observed by AMBER. The top and bottom panels display the high- and medium-resolution data, respectively. Error bars are calculated as the 1$\sigma$ noise level in the continuum. The array configurations, baseline lengths and position angles (N through E) are displayed in the top right corners of the panels. Colored squares in the bottom right corners connect properties of identical baselines.}
   \label{fig:data}
\end{figure*}
\clearpage

\addtocounter{figure}{-1}

\begin{figure*}[!h]
   \centering
   \includegraphics[height=0.85\textwidth, trim=0cm 0cm 0cm 13.5cm, clip=true, angle=90, origin=c]{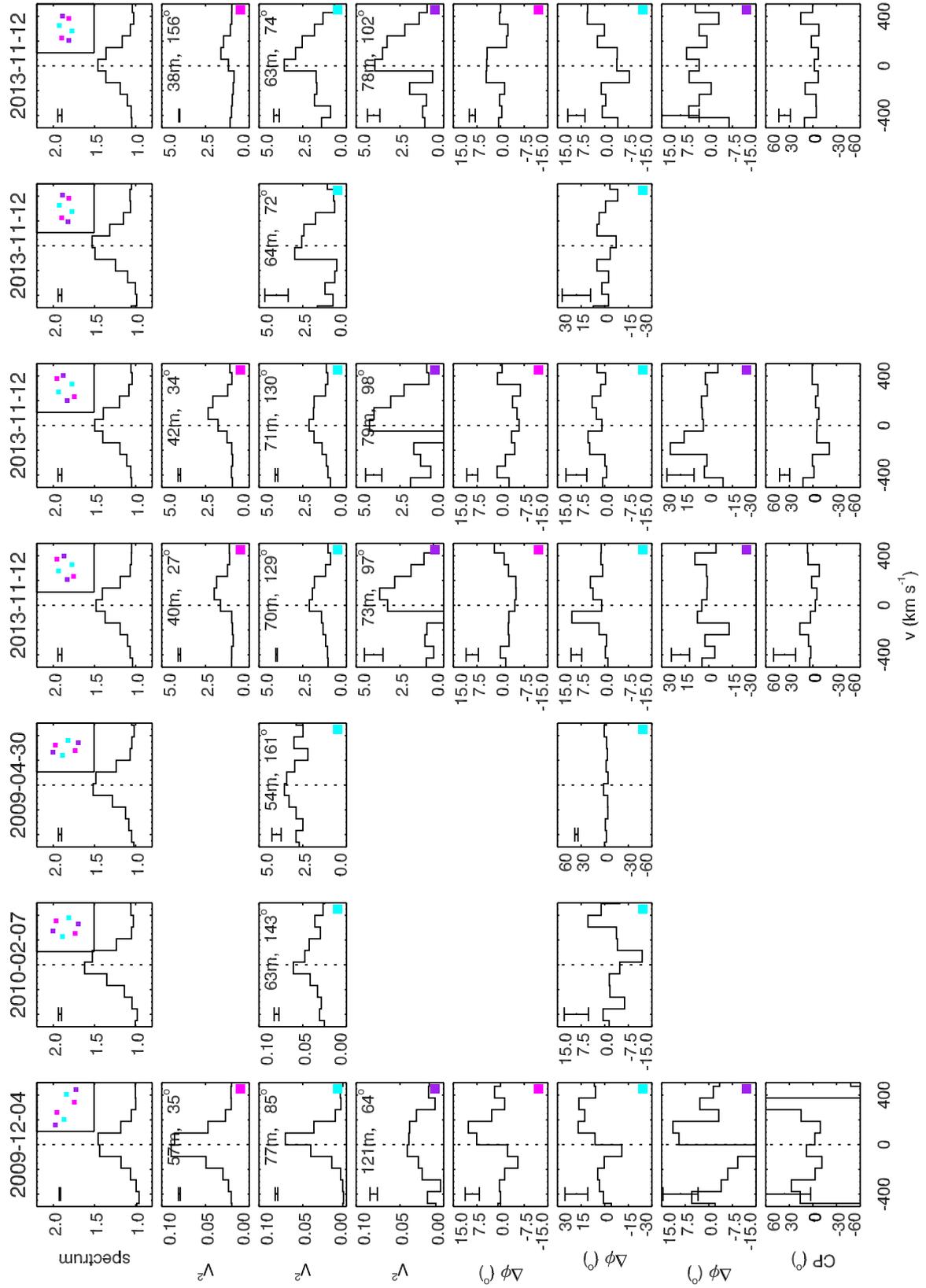}
   \caption{Continued.}
\end{figure*}
\clearpage

\subsection{Size scales}
\label{sec:results:size}

In this section we give an overview of the interferometric data. The AMBER data, which are centered on the Br$\gamma$ line, are displayed in Fig.~\ref{fig:data}. 
The error bars correspond to the root-mean-square variation in the continuum region at $500 < |\,\varv\,| < 1000$~km~s$^{-1}$. Two points are immediately apparent from the shape of the observed profiles. Firstly, the visibility is higher in the line than in the continuum for all baselines. This indicates that at the probed spatial scales, the Br$\gamma$ emitting region (hot gas) is more compact than the region producing the NIR continuum emission (mostly from dust). Secondly, in most baselines the regions emitting at a projected velocity are spatially resolved, or indicate a spatial shift in the photocenter, as indicated by the  ``wiggle" in the differential phases, across the Br$\gamma$ line. 

\begin{figure}[t]
   \centering
   \includegraphics[width=0.46\textwidth, page=1]{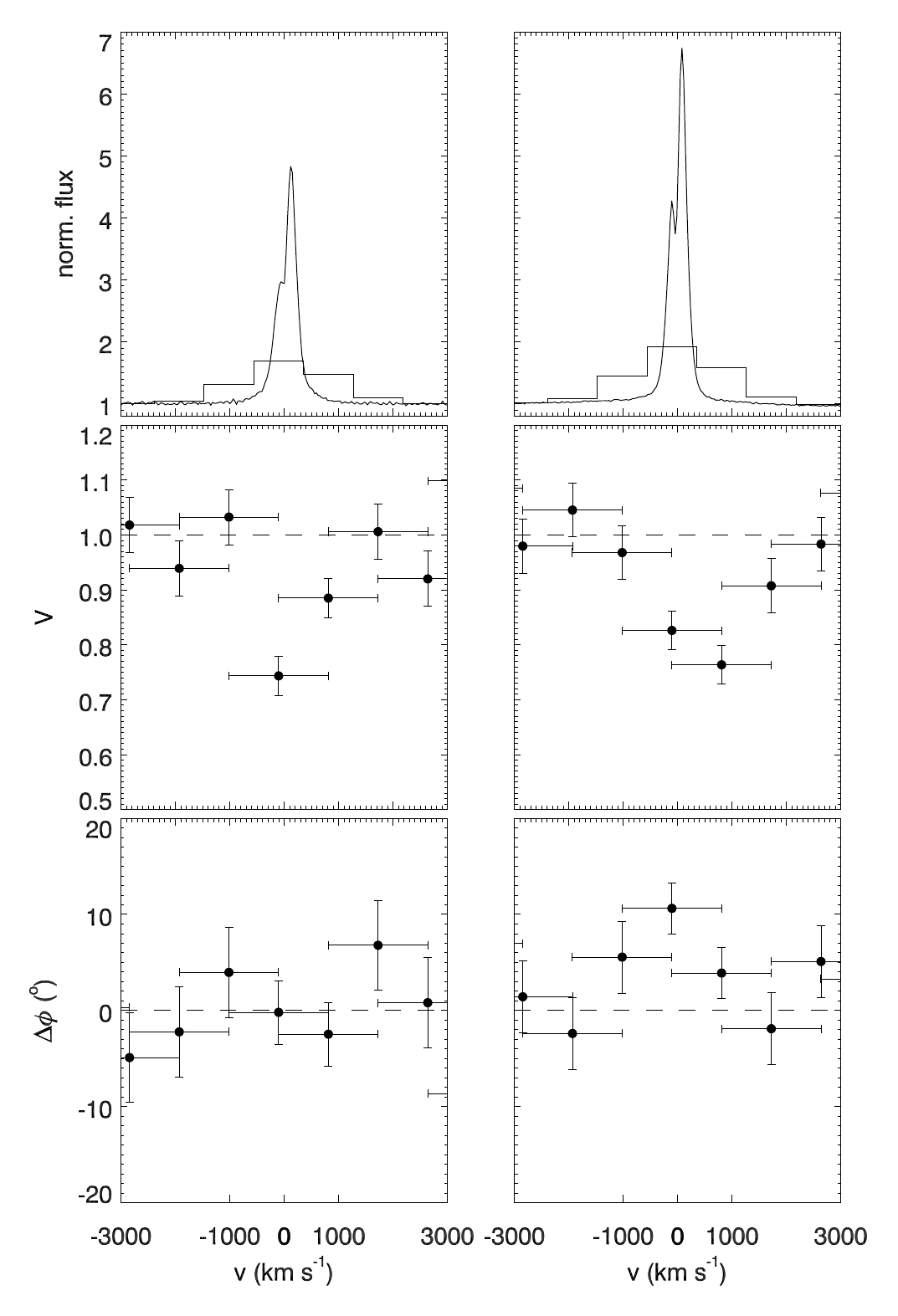}
   \caption{
   \textit{Top to bottom:} H$\alpha$ spectrum, squared visibility and differential phase taken with CHARA/VEGA, taken on 2010-10-11 (left panels) and 2012-10-29 (right panels). 
   }
   \label{fig:vega}
\end{figure}

\begin{table}[h]
  \centering
  \caption{\label{tab:sizeVEGA} Gaussian size estimates for the VEGA observations of H$\alpha$}
  \begin{tabular}{ccccc} 
     \hline  
     \hline
     Date & PA ($^\circ$) & $V$ & $V_{\rm line}$ & HWHM (mas) \\ 
     \hline
     2010-10-11 & 11 & $0.74\pm0.08$ & $0.37\pm0.20$ & $2.7\pm0.7$  \\
     2012-10-29 & --16 & $0.83\pm 0.04$ & $0.63\pm0.09$ & $1.8\pm0.3$   \\
     \hline 
  \end{tabular}
\end{table}

The VEGA data are displayed in Fig.~\ref{fig:vega}. The visibility drop across the H$\alpha$ line indicates that the H$\alpha$ emitting region is more extended than the star, which we assume to be the only contributor to  the continuum emission in the optical. The differential phases are, within the large error bars, consistent with zero. This indicates that the H$\alpha$ and continuum photocenters are close. 

\begin{figure}[t]
   \centering
   \includegraphics[width=0.46\textwidth, page=1]{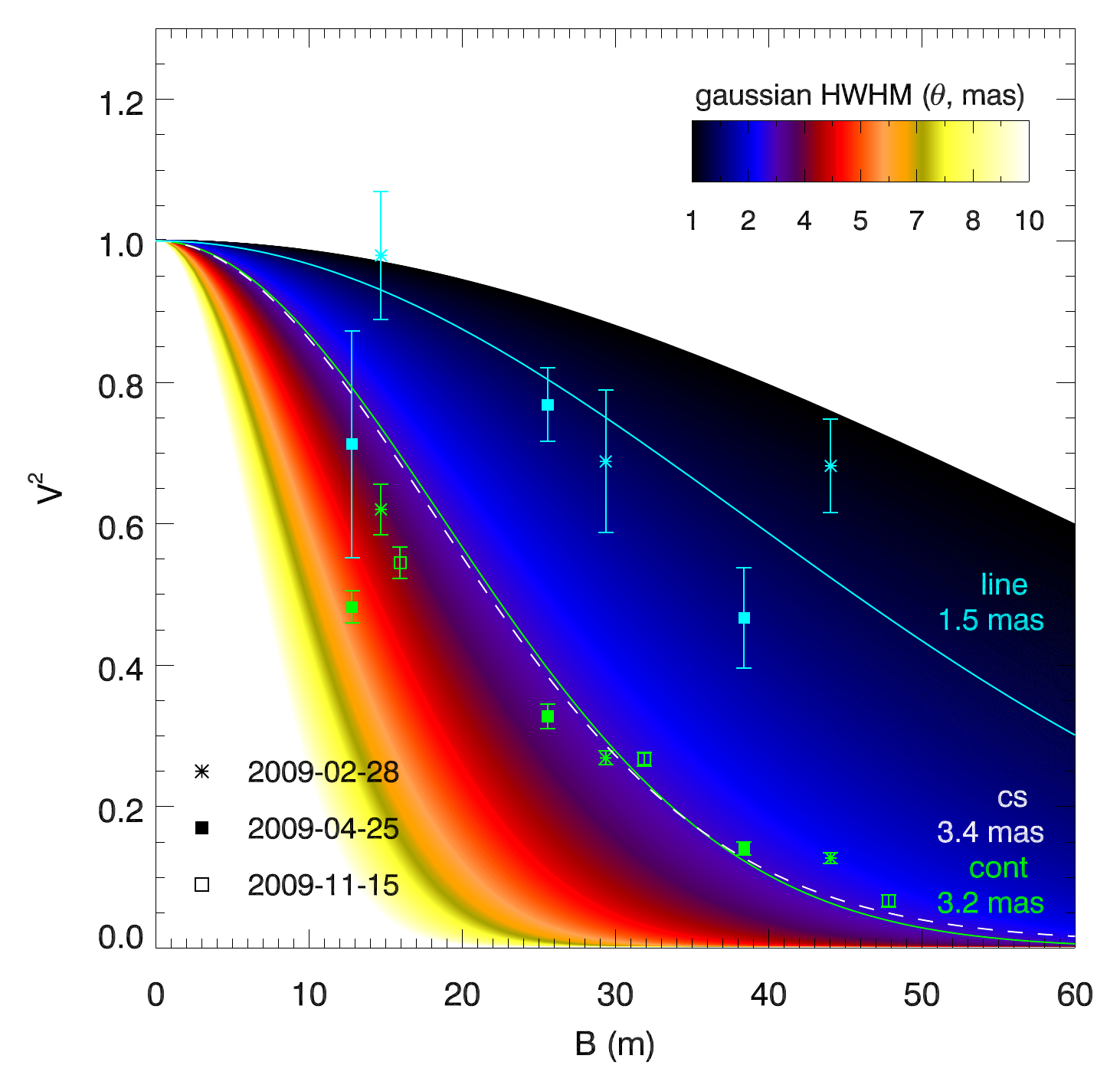}
   \caption{
 Average $V^2_{\rm cont}$, $V^2_{\rm line}$ values plotted as a function of baseline. The colored surface corresponds to the $V^2$ of a grid of single-Gaussian models with half width at half maximum (HWHM) as indicated by the colors. The best-fit models are indicated for the continuum-emitting region (green solid line) and the Br$\gamma$ line-emitting region ($\lambda_0 =2166.167$~nm, cyan solid line). The fit to the stellar-flux-corrected continuum region  ($V_{\rm cs}^2$) is also plotted (white dashed line).}
   \label{fig:size}
\end{figure}

We derive the size scales from the absolute visibilities. For VLTI observations carried out on the auxiliary telescopes, the absolute visibilities are well calibrated. We consider the three observations in the E0G0H0 configuration (1, 2 and 8, Table~\ref{tab:obs}). These baselines are the shortest available (hence the continuum is marginally resolved) and were aligned along PA~$=72^\circ-74^\circ$ which coincides with the disk major axis found by BF11. The VEGA baselines are close to PA~$=0^\circ$, more or less perpendicular to the disk major axis. 

In the NIR, we assume that the brightness distribution of the source in the spectral window of interest ($2159 < \lambda < 2173$~nm, $|\,\varv\,|<1000$~km~s$^{-1}$) is made up of three sources, (i) the continuum emission from the star, denoted by subscript ``$*$";  (ii) the continuum emission from the circumstellar material, ``cs" and (iii) the line emission from the circumstellar gas, ``line".  To disentangle the different components in the data we use the same approach as described in \citet[e.g.,][]{Weigelt2007}. This method uses observed visibilities and phases to separate out the contributions of the line- and continuum-emitting regions, assuming their relative flux contributions as described in Sect.~\ref{sec:modeling:configuration}.

In order to derive absolute sizes, we compare the visibilities of the components to those of a simple Gaussian intensity distribution. This assumes that the continuum and line brightness distributions are well approximated by a Gaussian profile, and are furthermore centro-symmetric and not variable. Fig.~\ref{fig:size} shows the observed visibilities over a grid of Gaussian models as a function of baseline.  The green symbols represent $V^2_{\rm cont}$, measured as the mean value of $V^2$ across the range $500 < |\,\varv\,| < 1000$~km~s$^{-1}$. For the high-resolution observations, $V^2_{\rm line}$ is computed and averaged (cyan symbols) over the central 8 wavelength bins ($ |\,\varv\,| < 50$~km~s$^{-1}$, or two resolution elements). In the medium-resolution observations, no meaningful values of $V^2_{\rm line}$ are retrieved, as the differential phase signal is very weak.   The best-fitting models for $V^2_{\rm cs}$ and $V^2_{\rm line}$ have a HWHM of $\theta_{\rm cs, mod}=3.4$~mas (1.7~au; $\chi^2_{\rm red}=38.1$) and $\theta_{\rm line, mod}=1.5$~mas (0.8~au; $\chi^2_{\rm red}=2.6$), respectively. The high $\chi^2$-value of the $V^2_{\rm cs}$-fit results from a bad fit of the short-baseline visibilities. This indicates a probable additional contribution from  an extended emission source (e.g. from an envelope or light scattered on outer disk layers), which could possibly be resolved by single-dish continuum observations. This model is corrected for the stellar flux and falls off to $f^2$ instead of zero. These best-fit values are used as fiducial size parameters along the disk major axis in the model presented in Sect.~\ref{sec:modeling}.

In the optical, the brightness distribution of the source is made up of only two components: the continuum emission from the star and the line emission from the circumstellar gas. The observed H$\alpha$ visibilities (Fig.~\ref{fig:vega}; Tab.~\ref{tab:sizeVEGA}) are converted to line visibilities $V_{\rm line}$ following the same approach as above. The visibilities correspond to a line-emitting region with Gaussian HWHM in the range $\theta_{\rm line, mod}$=1.5--3.4~mas ($\sim$0.8--1.7~au. The position angles of the VEGA baselines  have angles 60$^\circ$ and 87$^\circ$ to the disk major axis. The H$\alpha$-emitting region is thus probably slightly more extended than the Br$\gamma$-emitting region.

\begin{figure*}[!th]
   \centering
   \includegraphics[width=0.48\textwidth, page=1]{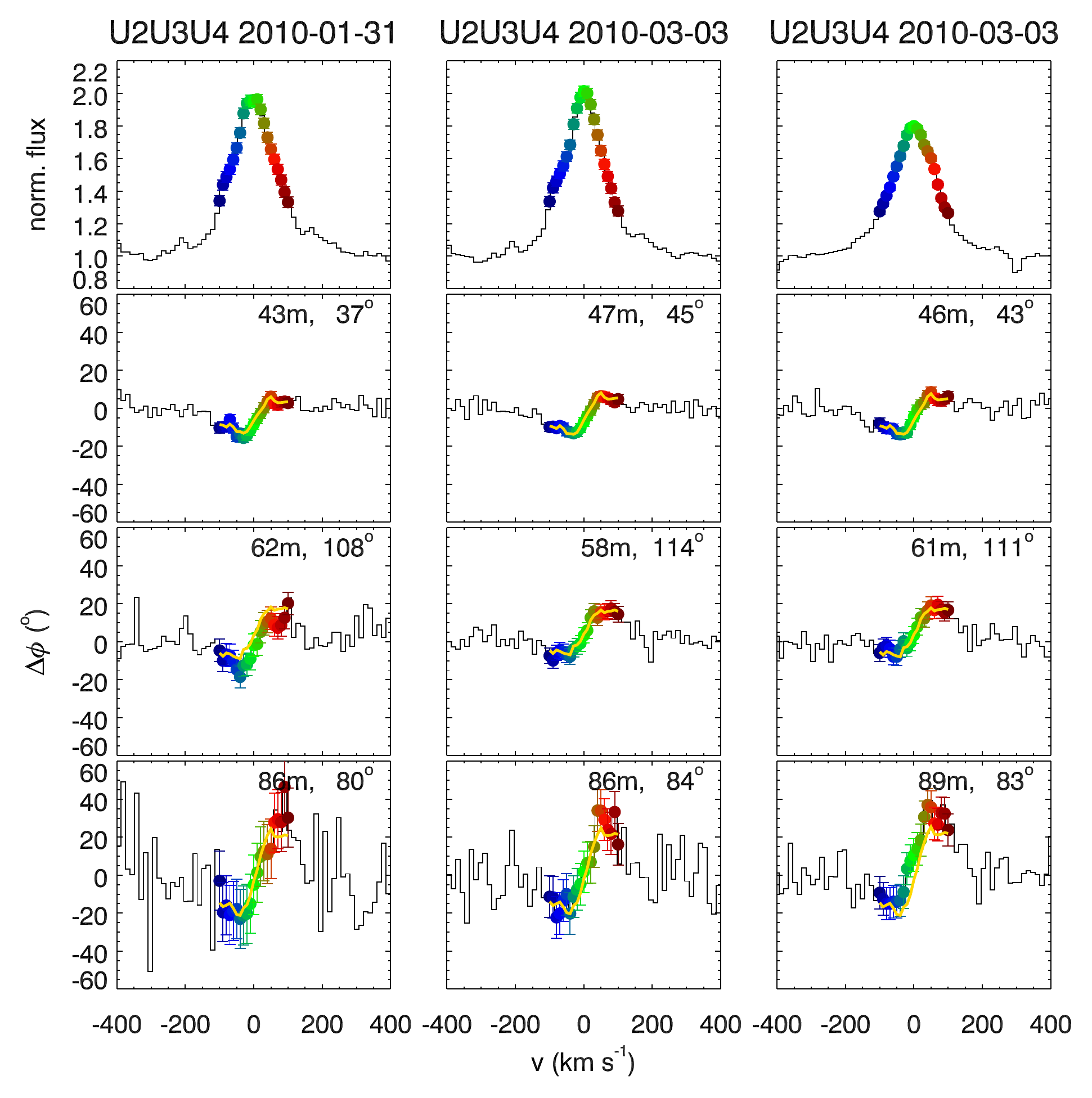}
   \includegraphics[width=0.46\textwidth, page=5]{astrometry.pdf}\\
   \includegraphics[width=0.21\textwidth, page=1]{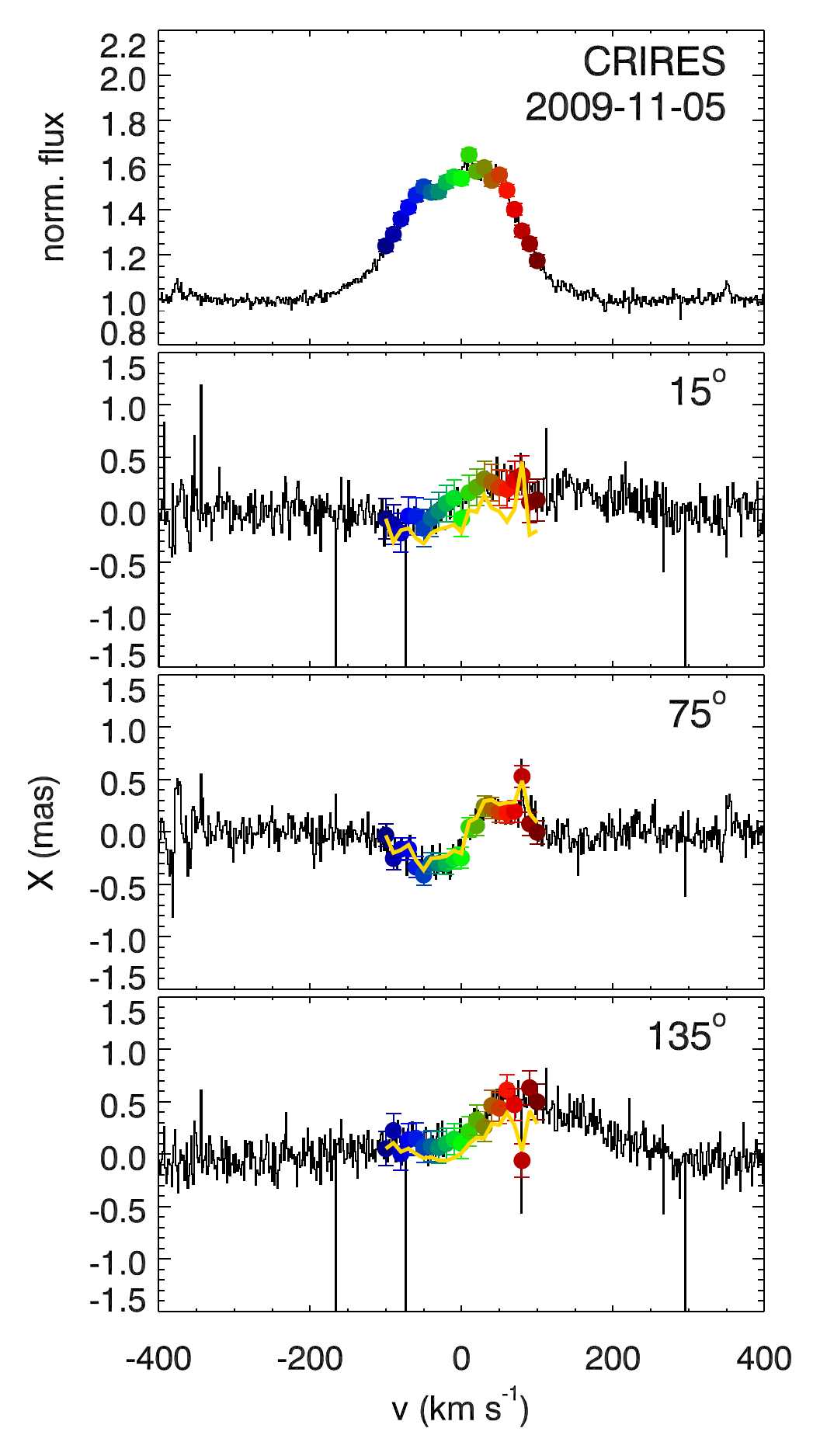}
   \includegraphics[width=0.38\textwidth, page=2]{astrocrires_bin.pdf}
   \includegraphics[width=0.38\textwidth, page=4]{astrocrires_bin.pdf}
     \caption{
\textit{Top left}: Br$\gamma$ spectrum and differential phase measured over the three U2U3U4 baselines in three different observations with AMBER. Overplotted are the data rebinned on $\Delta \varv=20$~km~s$^{-1}$ from -210 to 210~km~s$^{-1}$, color-coded with their velocity. The yellow lines correspond to the astrometric solution, $\vec{P}(\lambda)$, converted back to $\Delta\phi$. 
\textit{Top right}: two-dimensional representation of the photocenter displacement, $\vec{P}(\lambda)$, as a function of velocity across the Br$\gamma$ line. North is up, East is to the left. The (averaged) baselines of the U2U3U4 triplet are indicated on the bottom left. The dashed line corresponds to the orientation of the disk major axis ($\psi=71^\circ$) as derived by BF11.
\textit{Bottom left}: Spectrum and spectro-astrometric signal detected by CRIRES; same colors as above. 
\textit{Bottom middle}: photocenter displacement for CRIRES observations, continuum-corrected.
\textit{Bottom right}: photocenter displacement for CRIRES observations along the disk major axis ($\psi=71^\circ$).
      }
   \label{fig:astrometry}
\end{figure*}


\subsection{Photocenter shifts}
\label{sec:results:spectroastrometry}

The differential phase is related to the shift of the photocenter across the Br$\gamma$ line. It can be used to perform spectro-astrometry if the baselines sufficiently cover the two dimensions in the $(u,v)$-plane. We use the three U2U3U4 observations (3, 4 and 5, Table~\ref{tab:obs}), each with three baselines, which were taken with at most one month in between observations. These observations are selected to reduce the effect of systematical errors. Eventually, we check the solution for consistency with other observations. To convert the $\Delta\phi$-measurements to the photocenter displacement vector, $\vec{P} = \{P_{x}, P_{y}\}$, we follow \citet{Lachaume2003}. We set $\vec{P}=0$ in the continuum, and implicitly assume that $CP=0$ across the spectral range under scrutiny.

For marginally resolved objects (i.e., $V^2\gtrsim0.8$) and for small displacements ($B/\lambda \theta \ll 1$), the differential phase $\Delta\phi_i$ approximates a linear projection of $\vec{P}$ along the baseline vector $\vec{B}= -2\pi \, \{u_i,v_i\}/\lambda$. We will use this approximation even though our observations are not in the marginally resolved regime. We will comment below on how this is justified.

From the nine measurements $\vec{\Delta\phi} = \{ \Delta\phi_1, \Delta\phi_2, \dots, \Delta\phi_9 \}$ the displacement is obtained by performing a weighted linear least square fit: 
\begin{equation}
\vec{P} = (\vec{B}^{\rm T} \vec{W} \vec{B})^{-1}\, \vec{B}^{\rm T} \vec{W} \vec{\Delta\phi},
\end{equation}
where $\vec{W}$ is a $9\,\times\,9$ diagonal matrix containing the inverse squared errors on $\vec{\Delta\phi}$. We repeated this procedure for 21 velocity channels on the interval $\varv=(-210,210)$~km~s$^{-1}$, rebinned with a width of 20~km~s$^{-1}$. 

The result is plotted in Fig.~\ref{fig:astrometry} (top right). The line-emitting region is elongated NE to SW along PA~$\sim70^\circ$, with an overall offset towards the NW with respect to the continuum. The position angle agrees with the disk major axis derived by BF11, which is perpendicular to the polarization angle. The blue-shifted and red-shifted parts of the line are clearly separated between the disk hemispheres. The largest spatial offsets are seen in the lowest velocity channels. This result suggests that the Br$\gamma$ emission originates in a disk at $\psi \sim 71^\circ$ with a radially decreasing rotation around an axis of $\sim 160^\circ$. The result is consistent with the $\Delta\phi$ measurements from the other AMBER datasets which were not included in the fit (Fig.~\ref{fig:astromtetry:misc}, bottom), indicating that no additional asymmetries are resolved at this angular resolution. Other rotation fields, e.g. rigid rotation or equatorial outflow, are not suggested as the velocity decreases with radius and the velocity gradient is parallel to the major disk axis.

The Br$\gamma$ spectro-astrometric signal obtained with CRIRES (Fig.~\ref{fig:astrometry}, bottom) is consistent with the AMBER differential phases. The differentially rotating disk-like structure and the offset to the NW are both seen in these data. The consistency of the AMBER and CRIRES results, as well as the agreement with the AMBER observations which were not included in the solution, justify the use of the \citet{Lachaume2003} method, despite the low values of $V^2$.


\begin{figure*}[!th]
   \centering
   \includegraphics[width=\textwidth]{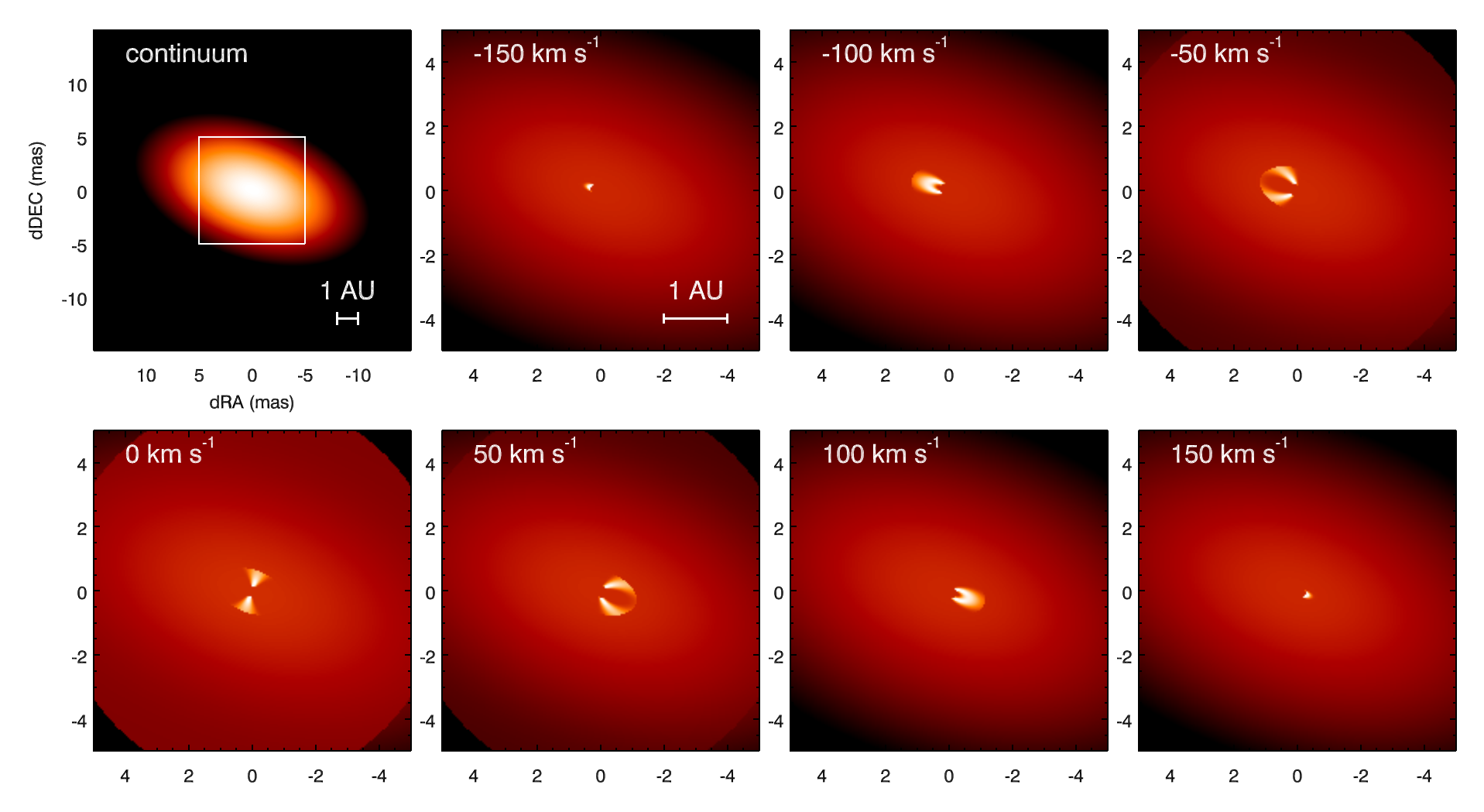}
   \caption{Selected channel maps of the geometric Keplerian disk plus spherical halo model (Model D0.6H; see text for description). The upper left panel shows the continuum map, the white box denotes the plot region for the other panels. The brightness scaling is logarithmic; in the zoomed panels, the continuum was scaled down arbitrarily to emphasize the shape of the line emission. The width of the channels is 10~km~s$^{-1}$. Note the emission from the 3~au halo in the fourth to sixth panels (visible as the red circle which extends up to a radius of $\sim 6''$).}
   \label{fig:models}
\end{figure*}

The solution $\vec{P}(\lambda)$ is a proxy for the first moment of the brightness map at wavelength $\lambda$. It is strictly a ``center of gravity" of the line and continuum contributions to the brightness map, weighted by their respective fluxes. To obtain information on the absolute size and dynamics of the line emitting region, the contribution of the continuum emission must be subtracted from the solution. However, given the low S/N in $V^2$, it is not feasible to derive channel-to-channel photocenter shifts in the line. Instead, we use the parameters derived in this section as an input for a geometric model (Sect.~\ref{sec:modeling}). 


\section{Geometric modeling}
\label{sec:modeling}

The results presented in Sect.~\ref{sec:results} suggest the presence of an au-scale rotating gas disk around HD~50138.  This interpretation is based on different subsets of the high-resolution AMBER-data: Br$\gamma$ size scale estimates are based on short-baseline observations (Sect.~\ref{sec:results:size}), while the spectro-astrometry is based on long-baseline observations in which the velocity field is well resolved (Sect.~\ref{sec:results:spectroastrometry}). In this section, we aim to interpret \textit{all} the interferometric data in terms of a single, self-consistent physical interpretation. To this end, we construct a geometric model, whose predictions are compared to the data. 


The model configuration is presented in Sect.~\ref{sec:modeling:configuration}. The number of parameters is kept as small as possible. Rather than obtaining the best-fit values by brute-force optimization, we choose a more intuitive approach. Most of the parameters are set to \textit{a priori} estimated values based on independent results. Three representative models are presented to illustrate the effect of adding or changing some of the parameters. We motivate these choices in Sect.~\ref{sec:modeling:parameters}. From this model, we calculate spectra, differential visibilities, and differential phases. We compare these to the data in Sect.~\ref{sec:modeling:compare}, and discuss the constraints on and the degeneracies of the model parameters. 

\subsection{Model configuration}
\label{sec:modeling:configuration}

Our model consists of four physical components:
\begin{itemize}
\item[(i)] a star; 
\item[(ii)] a continuum-emitting disk;
\item[(iii)] a line-emitting disk; 
\item[(iv)] a line-emitting spherical halo. 
\end{itemize}
The star (with mass $M_*$) is simulated as a point source at the grid origin at a distance $d$ and with a fixed fraction $f$ of the total flux in the modeled wavelength domain. For the circumstellar dust continuum emission, we adopt a single elliptical Gaussian distribution with as its parameters the HWHM along its major axis ($a_{\rm cont}$), its orientation ($\psi$), and inclination ($i$). Both star and dust disk are centered at the origin. The gas disk is translated into a Br$\gamma$ image in different velocity channels across the line. Directed by the results of Sect.~\ref{sec:results:spectroastrometry}, which indicate differential rotation, we assume that the gas is distributed in a geometrically thin rotating disk with a Keplerian velocity field \citep[see also, e.g.,][]{Eisner2010, Kraus2012}:
\begin{equation}
\varv(r)=\sqrt{\frac{GM_*}{r}}.
\end{equation}
The gas disk has inner and outer radii $R_{\rm in}$ and $R_{\rm out}$, and has the same inclination and orientation as the continuum-emitting disk. We consider emission from the disk in the Br$\gamma$ line at $\lambda_0=2166.167$~nm in the optically thin limit. The emission measure depends on the local temperature and surface density. We parametrize the resulting radial surface brightness profile by a simple power law: 
\begin{equation}
I(r)=I_0\, \left(\frac{r}{R_{\rm in}}\right)^{-\alpha},
\end{equation}
where we fit $I_0$ and $\alpha$ to match the observed Br$\gamma$ spectrum.

We have also investigated the effect of adding a spherical halo of radius $R_{\rm halo}$ to the line emission model, to explain the variable emission at the systemic velocity. The spectral line profile is uniform across the halo, and centered at the systemic velocity. In reality, the observed line profile would be a result of various physical mechanisms (e.g., Doppler, temperature or pressure broadening). We simply assume a Gaussian profile, and set its HWHM, $\Delta \varv_{\rm halo}$, and flux level by scaling it to fit the observed Br$\gamma$ spectrum.

Summarizing, the model has nine parameters ($d$, $\psi$, $i$, $a_{\rm cont}$, $R_{\rm in}$, $R_{\rm out}$, $\alpha$, $f$, and $M_*$). Two more ($R_{\rm halo}$ and $\Delta \varv_{\rm halo}$) are added when a spherical halo is included. 

The intensity at a physical radius $r$ is given by
\begin{equation}
I(r,\lambda) = \frac{B(r)}{\sqrt{2\pi\sigma}} \exp\left[{-\frac{(\lambda-(\varv(r)/c) \lambda_0)^2}{2\sigma^2}}\right],
\end{equation}
where $\sigma$ is:
\begin{equation}
\sigma = \frac{\lambda_0}{2\sqrt{2\ln2}\, \mathcal{R}}.
\end{equation}

\begin{table*}[t]
\centering
\caption{Model parameters}
\label{tab:modelparams}
\begin{tabular}{lllll}
\hline
\hline
Parameter & & Value & &Obtained from \\
\hline
\multicolumn{5}{c}{\textit{Geometry}} \\
\hline
$\psi$ (N through E) & &71$^\circ$ & & BF11, spectro-astrometry (Sect.~\ref{sec:results:spectroastrometry})\\
$i$ & & 56$^\circ$ & & BF11 \\
\hline
\multicolumn{5}{c}{\textit{Continuum model}} \\
\hline
$f$ &  & 0.08 & & SED (Sect.~\ref{sec:sed}) \\
$a_{\rm cont}$ &  & 1.7~au (3.4~mas) & & $V_{\rm cs}$ (Sect.~\ref{sec:results:size})\\
\hline
\multicolumn{5}{c}{\textit{Keplerian gas disk}} \\
\hline
$M_*$ &  & 6~M$_\odot$ & & BF09; this paper \\
$\alpha$ &  & $2$ & & \citet{Carciofi2008} \\
$R_{\rm in}$ &  & 0.1~au (0.2 mas)  & & Width of spectral line profile\\
& \textbf{D0.6} & \textbf{D3.0} & \textbf{D0.6H} & \\
$R_{\rm out}$ & 0.6~au (1.2~mas) & 3~au (6~mas) & 0.6~au (1.2~mas) & $V_{\rm line}$ (Sect.~\ref{sec:results:size}); peak-to-peak separation (Fig.~\ref{fig:xshooter}). \\
$R_{\rm halo}$ & -- & -- & 3~au (6~mas)  & \\
$\Delta\varv_{\rm halo}$ & -- & -- & 60~km~s$^{-1}$ & \\
\hline
\end{tabular}
\end{table*}

The contribution of every pixel is calculated with high spectral resolution ($\Delta \varv = 0.5$~km~s$^{-1}$). Channel maps are created by summing all contributions with a spectral bin width of $\Delta \varv = 10$~km~s$^{-1}$. A selection of channel maps is displayed in Fig.~\ref{fig:models}. Every channel map is superposed on the continuum image. The two contributions are scaled to match the observed line peak to continuum ratio in the spectrum, which is close to 2 for the high-resolution observations and close to 1.5 for the medium-resolution observations.

Simulated observables are obtained from the model images, as follows. The spectrum is calculated as the integrated flux of the individual channel maps. For every baseline used in the observations, complex visibilities were extracted from the line and continuum maps. These were converted to observed quantities $V^2$ and $\Delta \phi$ by taking the norm and argument of the Fourier transform at the (u,v) point corresponding to the observations. As the absolute calibration is uncertain, we compared squared differential visibilities $V^2_{\rm diff} = V^2 / V^2_{\rm cont}$ between models and observations.

\subsection{Model parameters}
\label{sec:modeling:parameters}


In this section we present the three sets of model parameters (see Tab.~\ref{tab:modelparams}) which we use to interpret the data. We introduce three representative models, which differ in their value of the outer disk radius, $R_{\rm out}$, and the presence of a gas halo. These representative models are: a 0.6~au gas disk (D0.6), a 3.0~au gas disk (D3.0), and a 0.6~au gas disk plus 3.0~au halo (D0.6H). The disk outer radius $R_{\rm out}$ varies between models D0.6 and D3.0. Model D0.6H is the same as D0.6, but with the additional parameters $R_{\rm halo}$ and $\Delta\varv_{\rm halo}$. The other eight model parameters are the same in all three cases. These are the geometric parameters $d$, $\psi$, $i$; the stellar parameters $f$ and $M_*$; the dust disk size, $a_{\rm cont}$; and the gas disk parameters $R_{\rm in}$ and $\alpha$. \textit{A priori} estimates are based on the spectroscopy results (Sect.~\ref{sec:results:spectroscopy}) and on previous studies. 

The adopted distance to the system is $d=500$~pc \citep{VanLeeuwen2007}. The dust continuum emission is modeled by an elliptical Gaussian. We adopt an orientation of the major axis, $\psi=71^\circ$, and an inclination angle, $i=56^\circ$, based on near- and mid-infrared continuum interferometry by BF11. The baselines used for the size estimates in Sect.~\ref{sec:results:size} were almost parallel to $\psi$. The best-fit value for the HWHM of the Gaussian fit to $V_{\rm cs}$ is 3.4~mas (1.7~au); we thus set $a_{\rm cont}$ to this value.

The parameter $f=0.08$ was obtained from SED fitting (Sect.~\ref{fig:sed}). The stellar mass is estimated by comparing $L_*$ and $T_{\rm eff}$ to pre- and post-main-sequence evolutionary models (BF09; see also \citealt{Schaller1992, Hosokawa2010}); we adopt $M_*=6$~M$_\odot$. The highest velocities in the line profile trace the disk inner radius. No emission is detected at velocities $| \, \varv\, |>230$~km~s$^{-1}$, hence we assume $R_{\rm in}= 0.1$~au. 

Two important physical features which are to be constrained by the model are the extent of the disk, and the presence of a halo. The (apparent) extent of the disk is determined by the combination of $\alpha$ and $R_{\rm out}$. We adopt $\alpha=2$, which matches the observed Br$\gamma$ spectrum, and is also a typical value for a (Herbig) Be star disk \citep{Carciofi2008, Eisner2010}. The results of Sect.~\ref{sec:results:size} yield a size of $\sim3.1$~mas for the line-emitting region along $\psi$, corresponding to a radius of 0.8~au. An independent estimate of the outer disk radius is derived from the double-peaked lines in the spectrum (Fig.~\ref{fig:xshooter}). If these are a result of rotation, the peak-to-peak separation corresponds to the disk diameter \citep{Horne1986}. The average separation of $\sim200$~km~s$^{-1}$ implies a disk outer radius of 0.5~au. Interpolating  between these two estimates, we adopt $R_{\rm out}=0.6$~au. A model with $R_{\rm out}=3$~au is also calculated to show the effect of the disk size on the observables. These models are referred to as D0.6 (the 0.6~au disk) and D3.0 (the 3~au disk).

Motivated by the variable emission at the systemic velocity, a third model, D0.6H, is constructed with the same parameters as D0.6, but with the addition of a spherical halo. Its radius, $R_{\rm halo}$, is set to 3~au, as this radius most accurately reproduces the observed $V^2_{\rm diff}$ values. The width of the spectral line profile is set to $\Delta\varv_{\rm halo}=60$~km~s$^{-1}$ to fit the observed emission.

\subsection{Comparing observations with models}
\label{sec:modeling:compare}

\begin{figure}[!t]
   \centering
   \includegraphics[width=0.99\columnwidth]{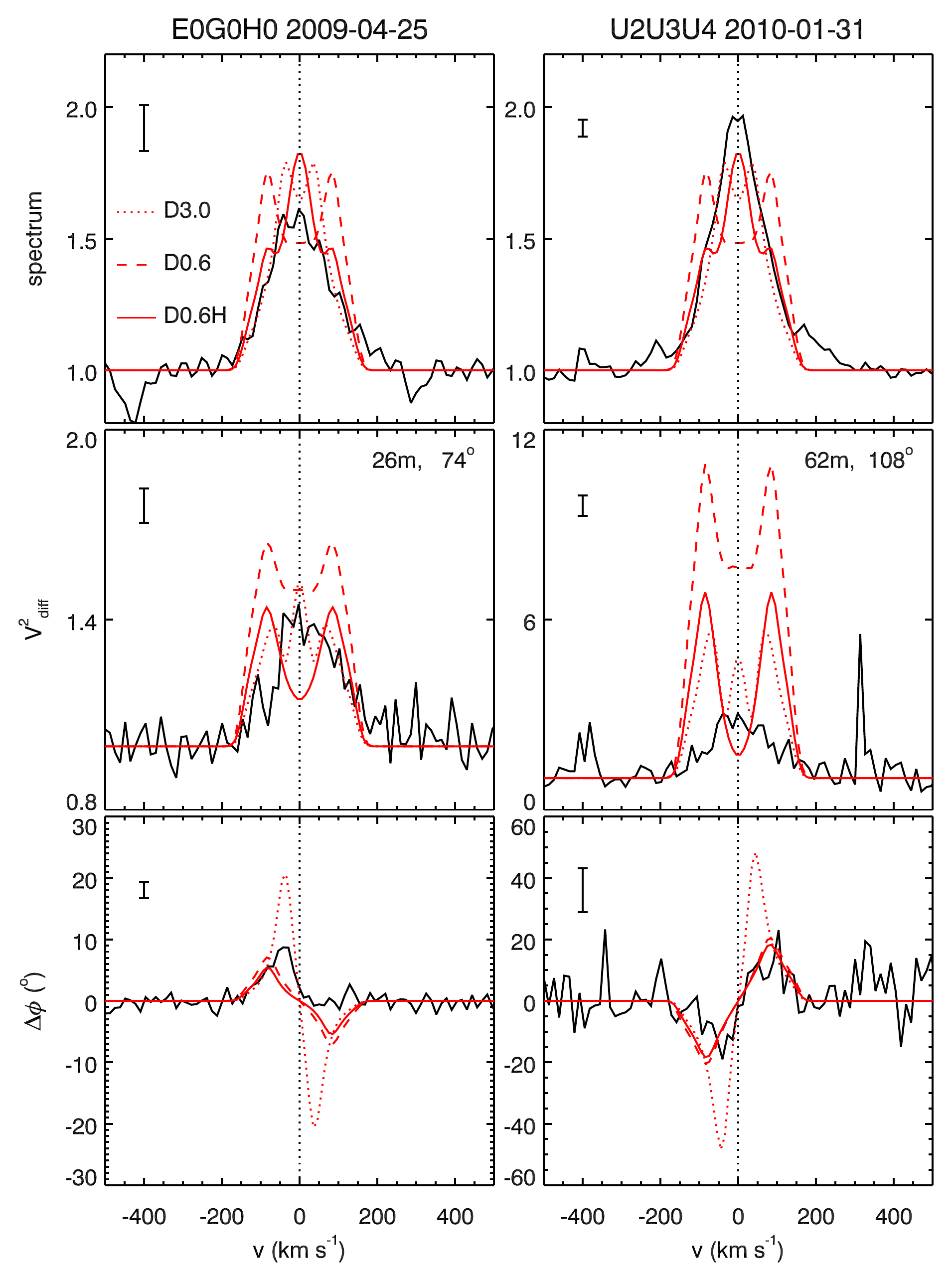}
   \caption{
   Two representative model fits. From top to bottom: spectrum, differential visibilities and differential phases observed with a compact (\textit{left}) and extended (\textit{right}) VLTI array. Overplotted are the observables calculated from models D0.6 (0.6~au disk, red line), D3.0 (3.0~au disk, blue line) and D0.6H (0.6~au disk + 3.0~au halo, green line). Fig.~\ref{fig:data_models_select} compares the results of the same models with all the AMBER observations.} 
   \label{fig:data_models_ut}
\end{figure}

In Fig.~\ref{fig:data_models_ut} two representative observations (one with short AT baselines, one with long UT baselines) are shown. The D0.6, D3.0, and D0.6H modeled spectra, visibilities and phases are overplotted. Fig.~\ref{fig:data_models_select} displays the same for all the observations. Model D0.6 correctly predicts the observed differential phase signatures (which was also suggested by Fig.~\ref{fig:astrometry}, bottom). However, it produces a double-peaked spectral line profile, which is not observed. Moreover, the predicted differential visibilities are too high, implying that the emission is too bright and compact with respect to the continuum. Model D3.0 (with larger outer gas disk radius) has a better fit to the spectra and visibilities, but a too high amplitude in the phase signal. 

This discrepancy may be summarized as follows: the phases indicate Keplerian rotation at a compact (sub-au) scale, but the visibilities indicate a significant fraction of the emission comes from larger (several au) scales. Furthermore, the spectra suggest that the emission is not solely due to a rotating disk. This motivates the inclusion of a halo: model D0.6H. The discrepancies are partly resolved by this model. The double peaks of the line are filled in; the differential phase signal of model D0.6 is retained, while the visibilities are scaled down to the levels of model D3.0. A mismatch between this model and the observations is the visibility drop at zero velocity, an effect of the halo's large size. The simple model does not reproduce the global offset of the emission in the NW direction.

Upon varying the eight fixed parameters, we conclude that these are reasonably well constrained. The effect of changing $\psi$, $i$ and $\alpha$ has been investigated; no other choice of these parameters leads to a better match with the observations. A degeneracy exists between $d$, $M_*$ and the size parameters in the model. For example, a lower adopted stellar mass has an effect similar to a shorter observing distance, or a larger disk inner radius. However, the observations are best reproduced with the initial set of values for $d$, $M_*$, and $R_{\rm in}$, which are consistent with the literature values (see Tab.~\ref{tab:modelparams}). The size parameters $R_{\rm out}, R_{\rm halo}$ following from these are consistent with the absolute size scales derived in Sect.~\ref{sec:results:size}.

The differential visibilities observed with baselines exceeding 50~m are overpredicted by all the models. This indicates that at these size scales, the disk is more extended than is assumed by the (oversimplified) model. Alternatively, the very low S/N in $V_{\rm cont}$ ($\sim$0) may lead to a systematically underpredicted $V_{\rm diff}$. The substructure within the visibility peak, which is predicted by the models, is not observed in any of the baselines. Rather, the observed visibilities peak at zero velocity; this may point to a compact source of emission, like a binary companion. This is also a possible explanation for the photocenter offset between line and continuum (see Sect.~\ref{sec:results:spectroastrometry}), which is not predicted by our axisymmetric model. This is further discussed in Sect.~\ref{sec:discussion:model}.  

Summarizing, the AMBER spectra, visibilities and phases are reasonably well reproduced by model D0.6H: a 0.6~au Keplerian disk and a 3~au halo. The parameters are constrained within a reasonable range, and consistent with independent observational results. However, the system shows signs of complexity that are not well constrained by the available observations (e.g., non-axisymmetry, variability) and that cannot be described with our simple model.

\section{Discussion}
\label{sec:discussion}

In this section, we discuss the main constraints obtained from the analysis of the data and the modeling. We first discuss the kinematics of the circumstellar gas as derived from our analysis and modeling. We also comment on the properties of the circumstellar dust disk. We then briefly review the signatures of binarity obtained from our and other observations. Finally, the implications of our results for the object's evolutionary state are discussed.

\subsection{Kinematics of the circumstellar gas}
\label{sec:discussion:model}

We find strong evidence for the presence of an au-scale Keplerian gas disk around HD~50138. The visibilities in the Br$\gamma$ line imply that most of the emission originates within 4~mas, which corresponds to 2~au (Sect.~\ref{sec:results:size}). This is consistent with the near-infrared disk size estimates by BF11, as well as the size of the H$\alpha$ emitting region. Dust at the sublimation temperature ($\sim 1500$~K) is a likely source of the $K$-band continuum emission. A picture where gas emission dominates within the dust sublimation radius is consistent with the observed $V^2_{\rm line}>V^2_{\rm cont}$ for all baselines. This is similar to what has been found in systems with a comparable dust SED \citep{Kraus2008b, Eisner2010}. 

The differential phases and CRIRES spectro-astrometry show a signature of rotation around an axis of $\sim160^\circ$ (Sect.~\ref{sec:results:spectroastrometry}). The decrease in velocity  with the distance to the center is consistent with a central mass of $\sim 6$~M$_\odot$ at a distance to the observer of 500~pc. Evidence for a gas disk with the same orientation was found in spectropolarimetry \citep{Bjorkman1998, Oudmaijer1999}. The inclination and orientation are consistent with the mid- and near-infrared continuum-emitting disk found by BF11. Similar evidence of a Keplerian velocity field has been found in interferometric studies of other B[e] and pre-main-sequence systems of comparable mass  \citep{Kraus2012, Wheelwright2012c}.


An additional, strongly variable emission peak at $\varv~\sim~0$ and with $\Delta\varv\sim60$~km~s$^{-1}$ is seen in some lines (e.g., Br$\gamma$, O~{\sc i}). These lines alternate between double- and single-peaked profiles on timescales of days to years. This suggests variable emission from the disk or the presence of an additional star in the system, although current campaigns have not found periodicity in the spectroscopic variability. The $\lesssim 1.7$~day period found in spectral lines by BF12 is attributed to stellar pulsations, reminiscent of pulsating late-type Be stars. Observations with temporal coverage at typical orbital timescales (i.e. weeks to months) are needed to constrain this possibility. We fit the emission component at the systemic velocity with a uniform line-emitting spherical halo, that could represent emission from a disk wind with a variable mass-loss rate. 

Aside from a disk and halo, other possible geometries include line emission from infall or outflow, but are less likely. Infall is suggested by several He\one, Mg~{\sc ii} and Si~{\sc ii} profiles, but the interpretation depends on the adopted systemic velocity. The spectral profile of the Br$\gamma$ line does not show absorption components, nor does its velocity distribution resemble an infall geometry. 

More commonly, however, this line traces outflow in a disk wind \citep{Malbet2007, Tatulli2007b, Kraus2008b, Benisty2010b, Weigelt2011}. The velocity is expected to increase as the material moves away from the source; in the case of HD~50138, a velocity decrease is observed. Also, asymmetries would exist between the red- and blue-shifted parts of the line emission, as the receding (red-shifted) jet lobe is blocked by the circumstellar disk \citep[see, e.g.,][]{Ellerbroek2013, Ellerbroek2014}. These are not observed. Finally, the outflow axis would be perpendicular to the polarization angle, which is not likely in a disk-jet geometry \citep{Maheswar2002}. The combination of a single-peaked profile and a spectro-astrometric rotation signature has been found in the CO lines of some protoplanetary disks \citep{Bast2011, Pontoppidan2011, Brown2013}. In these cases, the single peak is fitted by including an equatorial outflow component in the disk velocity field. Excretion disks of Be stars are not seen to have a significant outward velocity component \citep{Oudmaijer2011, Meilland2012, Wheelwright2012b}.



Another important result is the offset of the Br$\gamma$ line emission towards the NW with respect to the continuum photocenter (Fig.~\ref{fig:astrometry}, top right). This may indicate that the line-emission disk geometry is more complicated; for example, a flaring inner disk \citep[equivalent to the outer disk geometry described in][]{Lagage2006} would naturally induce a photocenter offset. Alternative scenarios include an asymmetric inner disk structure or a close companion. An asymmetric distribution of the continuum emission would also influence the offset. The offset is in the polar direction, which would be a natural consequence of a disk rim; in the case of a binary this offset direction would be coincidental. Although the spatial extent of the Gaussian considered for the continuum is consistent with the estimated location of R$_{\rm sub}$ (Eq. 1 in \citealt{Dullemond2010};  $1.5-4$~au (3-8 mas)),  a more complicated spatial structure than a Gaussian is expected for the near-infrared emission.  A Keplerian gas disk may exist in case of a close binary scenario as a result of mass exchange. Further indications of possible binarity are the ambiguous spectral classification and variability, although pulsations may cause the latter (BF09, BF12). 

The photocenter offset in the H$\alpha$ line found by \citet{Baines2006} is in the same NW direction as the Br$\gamma$ shift. In this part of the spectrum, however, the continuum is dominated by photospheric emission; therefore, these authors interpret the offset as a signature of a (wide) binary. A more thorough investigation of the near-infrared continuum visibilities and phases will contribute to resolving possible companion(s) or an asymmetric disk (Kluska et al., in prep.). Close companions may also be found or excluded by a spectroscopic monitoring campaign. 

To summarize, most observed signatures are consistent with the presence of an au-scale Keplerian disk plus gaseous halo. Signatures which are not accounted for are the absolute level of the long-baseline visibilities, the variability of the emission and the  photocenter offset. Additional sources of emission (e.g. a binary companion) and/or a more complicated disk geometry are required to obtain a satisfactory explanation for all the data.

\subsection{Evolutionary state}
\label{sec:discussion:evolutionary}

The nature of HD~50138 is not clear: is it a pre-main-sequence (i.e, Herbig B[e] star), main-sequence or post-main-sequence object? Most of the observed characteristics are consistent with all of these possibilities. In this section we will briefly review these arguments.

We find that the Br$\gamma$-emitting circumstellar gas has a rotation-dominated velocity field, most likely a disk. au-scale Keplerian gas disks are found around both pre-main-sequence \citep{Acke2005, Bagnoli2010, Weigelt2011, Kraus2012} and post-main-sequence \citep{Bujarrabal2007, Wheelwright2012c} early-type stars, some of which are known binaries. In pre-main-sequence stars, accretion columns and outflows more commonly dominate the Br$\gamma$ emission \citep{Eisner2010}. The decretion disks of post-main-sequence systems are expected to have a strong outflow component \citep{Lamers1991}. In the case of HD~50138, rotation dominates the velocity field, which is consistent with both a pre- and post-main-sequence nature.


The existence of a close binary companion, as suggested by various signatures (photocenter offset, ambiguous spectral type, variability), is also consistent with both scenarios. Close binary systems surrounded by dust disks are commonly seen in Herbig systems \citep{Baines2006, Wheelwright2011, Garcia2013} and post-AGB systems \citep{DeRuyter2006, Gielen2008, Kraus2013}. In the former case, the close binary may have formed by disk fragmentation \citep{Krumholz2009}. In the latter case, the dust disk is most likely a result of binary interaction, which would also give rise to shell phases. No signature of the companion star is seen in the spectral lines, which would indicate a low mass with respect to the primary. The SED does not enable us to distinguish between these scenarios. Thermal pulses and thus a post-AGB phase are excluded given the system's luminosity. Alternatively, the system could be a failed AGB star evolving into a sub-dwarf OB star \citep{Heber2009}. 

In other B[e] systems, high spatial and spectral resolution observations have yielded evidence of binarity and gas disks. Some of these are confirmed evolved systems, whose circumstellar gas and dust disk is a likely result of the interaction between the stellar companions (HD~87643, \citealt{Millour2009};, MWC~300, \citealt{Wang2012};, HD~327083, \citealt{Wheelwright2012c}). In a few cases, unambiguous evidence exists for an evolved evolutionary state (e.g., the $^{13}$CO abundance in HD~327083, \citealt{Wheelwright2012a}). Properties considered by some authors to be evidence of a pre-main-sequence nature are e.g., the absence of a binary companion \citep[HD~85567][]{Wheelwright2013} or the absence of an outflow component in the disk \citep[V~921 Sco][]{Kraus2012}.

The emission component in Br$\gamma$ and other spectral lines at $\varv\sim0$ appears and disappears, indicating a highly variable circumstellar geometry. Apart from our interpretation as a gaseous halo in Sect.~\ref{sec:modeling}, the emission may arise as intrinsic emission from a companion star, from mass transfer columns in a close binary, or from slow-moving material in the accretion region close to the star. This component, as well as the many other detected variable signatures, may relate to shell phases and outbursts. These are also seen in both post- and pre-main-sequence objects (\citealt{Crause2003}, \citealt{Kospal2013}, \citealt{Ellerbroek2014}). Peculiar spectral profiles (He~{\sc i}, Mg~{\sc ii}, Si~{\sc ii}) indicate either infall or outflow, and are thus inconclusive regarding the evolutionary state. The spectral variability is poorly constrained because of the limited time coverage of the observations. 



Additional heuristic arguments prefer a post-main-sequence nature. The observed emission from high Paschen, Brackett and Pfund transitions are less commonly seen in young stars \citep{Jaschek1998, Lamers1998}. The 2.3~$\mu$m CO emission is a common feature of young stellar objects in this mass range (\citealt{BikThi2004, Ochsendorf2011, Ilee2013, Ellerbroek2013, Ellerbroek2013b}), but is not observed in HD~50138. No clear signatures of accretion are seen, and the location of the object in the Hertzsprung-Russell diagram is inconclusive regarding its evolutionary state (BF09). Finally, the apparent isolation of the object from a star-forming region does not suggest a young age. Proper motion and distance measurements from the Gaia mission \citep{Perryman2001} will better constrain the formation history. 






In summary, in the case of HD~50138, a post-main-sequence and interacting binary nature is slightly favored over a pre-main-sequence nature, although many signatures are inconclusive. Given the extremely dynamical circumstellar environment, high-cadence spectroscopic and interferometric monitoring campaigns on timescales from days to years are the most promising strategy to further unravel this system. 

 \section{Summary and Conclusions}
 \label{sec:conclusion}
 
We have presented observations of the kinematics of the gaseous circumstellar environment of HD~50138. Our main conclusions are listed below:
\begin{itemize}
\item Strong evidence is found that the Br$\gamma$ emitting gas is distributed in a Keplerian rotating disk. This is suggested by the rotating and radially decreasing velocity field of the gas, which is distributed in an elongated structure aligned with independent estimates of the disk major axis.
\item The gas line emission originates from a smaller region (up to 3~au) than the continuum emission attributed to dust. This is consistent with the inner few au of the disk being too hot for dust to exist in equilibrium; in this region, the gas disk is expected to dominate the energy output.
\item The interferometric observables can qualitatively be reproduced with a model of a geometrically thin Keplerian disk surrounded by a low-velocity halo and a more extended source of continuum emission. Supporting evidence for the existence of these components is given by the spectrum of the source. 
\item The strong variability of shell- or disk-dominated spectral line profiles indicates that significant changes take place in the system's appearance on timescales as short as months, probably inhibiting a unified ``fit" to all the datasets.
\item The absolute offset of the photocenter in the continuum may be caused by an asymmetric disk geometry which affects the line- and continuum-emitting regions differently (e.g., a puffed-up rim seen under an inclination angle). Alternatively, a binary companion may cause the offset.
\item No definitive conclusion on the evolutionary state could be reached. The system is possibly a binary, and bears much resemblance to both Herbig B[e] and post-main-sequence systems. In view of mostly circumstantial evidence, the latter scenario is slightly favored.
\end{itemize}

After nearly a century of intensive research, HD~50138 continues to be an enigmatic object. In this study, we have for the first time mapped its inner environment and have discovered a rotating gas disk. Observations at a higher temporal resolution are key towards a better understanding of the evolutionary state of this highly dynamical system. The combined forces of interferometry and spectroscopy on high spatial and spectral resolution, and a broad wavelength domain, proves to be a very insightful strategy. Complementary observations at high spatial and spectral resolution of different line and dust tracers (e.g. taken with the Atacama Large Millimeter Array) will help resolving these elusive objects.

\acknowledgements
The authors thank Olga Hartoog for providing the FASTWIND models. Jerome Bouvier, Alex Carciofi, Ewine van Dishoeck, Carsten Dominik, Alex de Koter, Henny Lamers, Ren\'{e} Oudmaijer, Philippe Stee and Rens Waters are acknowledged for constructive discussions about the source. The ESO staff are acknowledged for technical support of the observations. The authors have made use of the \texttt{AMBER data reduction package} of the Jean-Marie Mariotti Center\footnote{Available at \texttt{http://www.jmmc.fr/amberdrs}}. We also used the \texttt{SearchCal} service \footnote{Available at \texttt{http://www.jmmc.fr/searchcal}} co-developped by Lagrange and IPAG, and of CDS Astronomical Databases SIMBAD and VIZIER \footnote{Available at \texttt{http://cdsweb.u-strasbg.fr/}}. The CHARA Array, operated by Georgia State University through the College of Arts and Sciences and NSF AST 12-11129, was built with funding provided by the National Science Foundation, Georgia State University, the W. M. Keck Foundation, and the David and Lucile Packard Foundation. LE and MB acknowledge a grant from the Fizeau Program, funded by WP14  OPTICON/FP7. ADS and MBF thank the CNRS-PICS program 2010-2012 for partial financial support.


\bibliographystyle{aa}

\begin{thebibliography}{98}
\expandafter\ifx\csname natexlab\endcsname\relax\def\natexlab#1{#1}\fi

\bibitem[{{Acke} {et~al.}(2005){Acke}, {van den Ancker}, \&
  {Dullemond}}]{Acke2005}
{Acke}, B., {van den Ancker}, M.~E., \& {Dullemond}, C.~P. 2005, \aap, 436, 209

\bibitem[{{Allen} \& {Swings}(1976)}]{Allen1976}
{Allen}, D.~A. \& {Swings}, J.~P. 1976, \aap, 47, 293

\bibitem[{{Andrillat} \& {Houziaux}(1991)}]{Andrillat1991}
{Andrillat}, Y. \& {Houziaux}, L. 1991, \iaucirc, 5164, 3

\bibitem[{Bagnoli {et~al.}(2010)Bagnoli, van Lieshout, Waters, van~der Plas,
  Acke, van Winckel, Raskin, \& Meerburg}]{Bagnoli2010}
Bagnoli, T., van Lieshout, R., Waters, L. B. F.~M., {et~al.} 2010, The
  Astrophysical Journal Letters, 724, L5

\bibitem[{{Baines} {et~al.}(2006){Baines}, {Oudmaijer}, {Porter}, \&
  {Pozzo}}]{Baines2006}
{Baines}, D., {Oudmaijer}, R.~D., {Porter}, J.~M., \& {Pozzo}, M. 2006, \mnras,
  367, 737

\bibitem[{{Bast} {et~al.}(2011){Bast}, {Brown}, {Herczeg}, {van Dishoeck}, \&
  {Pontoppidan}}]{Bast2011}
{Bast}, J.~E., {Brown}, J.~M., {Herczeg}, G.~J., {van Dishoeck}, E.~F., \&
  {Pontoppidan}, K.~M. 2011, \aap, 527, A119

\bibitem[{{Benisty} {et~al.}(2010){Benisty}, {{\mockalph{bbbb}}Malbet},
  {Dougados}, {Natta}, {Le Bouquin}, {Massi}, {Bonnefoy}, {Bouvier}, {Chauvin},
  {Chesneau}, {Garcia}, {Grankin}, {Isella}, {Ratzka}, {Tatulli}, {Testi},
  {Weigelt}, \& {Whelan}}]{Benisty2010b}
{Benisty}, M., {{\mockalph{bbbb}}Malbet}, F., {Dougados}, C., {et~al.} 2010,
  \aap, 517, L3

\bibitem[{{Bik} \& {Thi}(2004)}]{BikThi2004}
{Bik}, A. \& {Thi}, W.~F. 2004, \aap, 427, L13

\bibitem[{{Bjorkman} {et~al.}(1998){Bjorkman}, {Miroshnichenko}, {Bjorkman},
  {Meade}, {Babler}, {Code}, {Anderson}, {Fox}, {Johnson}, {Weitenbeck},
  {Zellner}, \& {Lupie}}]{Bjorkman1998}
{Bjorkman}, K.~S., {Miroshnichenko}, A.~S., {Bjorkman}, J.~E., {et~al.} 1998,
  \apj, 509, 904

\bibitem[{{Bopp}(1993)}]{Bopp1993}
{Bopp}, B.~W. 1993, Information Bulletin on Variable Stars, 3834, 1

\bibitem[{{Borges Fernandes} {et~al.}(2009){Borges Fernandes}, {Kraus},
  {Chesneau}, {Domiciano de Souza}, {de Ara{\'u}jo}, {Stee}, \&
  {Meilland}}]{Borges2009}
{Borges Fernandes}, M., {Kraus}, M., {Chesneau}, O., {et~al.} 2009, \aap, 508,
  309 (BF09)

\bibitem[{{Borges Fernandes} {et~al.}(2012){Borges Fernandes}, {Kraus},
  {Nickeler}, {De Cat}, {Lampens}, {Pereira}, \& {Oksala}}]{Borges2012}
{Borges Fernandes}, M., {Kraus}, M., {Nickeler}, D.~H., {et~al.} 2012, \aap,
  548, A13 (BF12)

\bibitem[{{Borges Fernandes} {et~al.}(2011){Borges Fernandes}, {Meilland},
  {Bendjoya}, {Domiciano de Souza}, {Niccolini}, {Chesneau}, {Millour},
  {Spang}, {Stee}, \& {Kraus}}]{Borges2011}
{Borges Fernandes}, M., {Meilland}, A., {Bendjoya}, P., {et~al.} 2011, \aap,
  528, A20 (BF11)

\bibitem[{{Bourg\'{e}s} {et~al.}(2014){Bourg\'{e}s}, {Lafrasse}, {Mella},
  {Chesneau}, {Bouquin}, {Duvert}, {Chelli}, \& {Delfosse}}]{Bourges2014}
{Bourg\'{e}s}, L., {Lafrasse}, S., {Mella}, G., {et~al.} 2014, in Astronomical
  Society of the Pacific Conference Series, Vol. 485, Astronomical Society of
  the Pacific Conference Series, ed. N.~{Manset} \& P.~{Forshay}, 223

\bibitem[{{Brown} {et~al.}(2013){Brown}, {Troutman}, \& {Gibb}}]{Brown2013}
{Brown}, L.~R., {Troutman}, M.~R., \& {Gibb}, E.~L. 2013, \apjl, 770, L14

\bibitem[{{Bujarrabal} {et~al.}(2007){Bujarrabal}, {van Winckel}, {Neri},
  {Alcolea}, {Castro-Carrizo}, \& {Deroo}}]{Bujarrabal2007}
{Bujarrabal}, V., {van Winckel}, H., {Neri}, R., {et~al.} 2007, \aap, 468, L45

\bibitem[{{Carciofi} \& {Bjorkman}(2008)}]{Carciofi2008}
{Carciofi}, A.~C. \& {Bjorkman}, J.~E. 2008, \apj, 684, 1374

\bibitem[{{Cardelli} {et~al.}(1989){Cardelli}, {Clayton}, \&
  {Mathis}}]{Cardelli1989}
{Cardelli}, J.~A., {Clayton}, G.~C., \& {Mathis}, J.~S. 1989, \apj, 345, 245

\bibitem[{{Chelli} {et~al.}(2009){Chelli}, {Utrera}, \& {Duvert}}]{chelli09}
{Chelli}, A., {Utrera}, O.~H., \& {Duvert}, G. 2009, \aap, 502, 705

\bibitem[{{Cidale} {et~al.}(2001){Cidale}, {Zorec}, \&
  {Tringaniello}}]{Cidale2001}
{Cidale}, L., {Zorec}, J., \& {Tringaniello}, L. 2001, \aap, 368, 160

\bibitem[{{Corporon} \& {Lagrange}(1999)}]{Corporon1999}
{Corporon}, P. \& {Lagrange}, A.-M. 1999, \aaps, 136, 429

\bibitem[{{Crause} {et~al.}(2003){Crause}, {Lawson}, {Kilkenny}, {van Wyk},
  {Marang}, \& {Jones}}]{Crause2003}
{Crause}, L.~A., {Lawson}, W.~A., {Kilkenny}, D., {et~al.} 2003, \mnras, 341,
  785

\bibitem[{{de Ruyter} {et~al.}(2006){de Ruyter}, {van Winckel}, {Maas}, {Lloyd
  Evans}, {Waters}, \& {Dejonghe}}]{DeRuyter2006}
{de Ruyter}, S., {van Winckel}, H., {Maas}, T., {et~al.} 2006, \aap, 448, 641

\bibitem[{{Doazan}(1965)}]{Doazan1965}
{Doazan}, V. 1965, Annales d'Astrophysique, 28, 1

\bibitem[{{Domiciano de Souza} {et~al.}(2007){Domiciano de Souza}, {Driebe},
  {Chesneau}, {Hofmann}, {Kraus}, {Miroshnichenko}, {Ohnaka}, {Petrov},
  {Preisbisch}, {Stee}, {Weigelt}, {Lisi}, {Malbet}, \&
  {Richichi}}]{Domiciano2007}
{Domiciano de Souza}, A., {Driebe}, T., {Chesneau}, O., {et~al.} 2007, \aap,
  464, 81

\bibitem[{{Dullemond} \& {Monnier}(2010)}]{Dullemond2010}
{Dullemond}, C.~P. \& {Monnier}, J.~D. 2010, \araa, 48, 205

\bibitem[{{Eisner} {et~al.}(2010){Eisner}, {Monnier}, {Woillez}, {Akeson},
  {Millan-Gabet}, {Graham}, {Hillenbrand}, {Pott}, {Ragland}, \&
  {Wizinowich}}]{Eisner2010}
{Eisner}, J.~A., {Monnier}, J.~D., {Woillez}, J., {et~al.} 2010, \apj, 718, 774

\bibitem[{{Ellerbroek} {et~al.}(2013{\natexlab{a}}){Ellerbroek}, {Bik},
  {Kaper}, {Maaskant}, {Paalvast}, {Tramper}, {Sana}, {Waters}, \&
  {Balog}}]{Ellerbroek2013b}
{Ellerbroek}, L.~E., {Bik}, A., {Kaper}, L., {et~al.} 2013{\natexlab{a}}, \aap,
  558, A102

\bibitem[{{Ellerbroek} {et~al.}(2014){Ellerbroek}, {Podio}, {Dougados},
  {Cabrit}, {Sitko}, {Sana}, {Kaper}, {de Koter}, {Klaassen}, {Mulders},
  {Mendigutia}, {Grady}, {Grankin}, {van Winckel}, {Bacciotti}, {Russell},
  {Lynch}, {Hammel}, {Beerman}, {Day}, {Huelsman}, {Werren}, {Henden}, \&
  {Grindlay}}]{Ellerbroek2014}
{Ellerbroek}, L.~E., {Podio}, L., {Dougados}, C., {et~al.} 2014, ArXiv e-prints

\bibitem[{{Ellerbroek} {et~al.}(2013{\natexlab{b}}){Ellerbroek}, {Podio},
  {Kaper}, {Sana}, {Huppenkothen}, {de Koter}, \& {Monaco}}]{Ellerbroek2013}
{Ellerbroek}, L.~E., {Podio}, L., {Kaper}, L., {et~al.} 2013{\natexlab{b}},
  \aap, 551, A5

\bibitem[{{Garcia} {et~al.}(2013){Garcia}, {Benisty}, {Dougados}, {Bacciotti},
  {Clausse}, {Massi}, {M{\'e}rand}, {Petrov}, \& {Weigelt}}]{Garcia2013}
{Garcia}, P.~J.~V., {Benisty}, M., {Dougados}, C., {et~al.} 2013, \mnras, 430,
  1839

\bibitem[{{Gielen} {et~al.}(2008){Gielen}, {van Winckel}, {Min}, {Waters}, \&
  {Lloyd Evans}}]{Gielen2008}
{Gielen}, C., {van Winckel}, H., {Min}, M., {Waters}, L.~B.~F.~M., \& {Lloyd
  Evans}, T. 2008, \aap, 490, 725

\bibitem[{{Grady} {et~al.}(1996){Grady}, {Perez}, {Talavera}, {Bjorkman}, {de
  Winter}, {The}, {Molster}, {van den Ancker}, {Sitko}, {Morrison}, {Beaver},
  {McCollum}, \& {Castelaz}}]{Grady1996}
{Grady}, C.~A., {Perez}, M.~R., {Talavera}, A., {et~al.} 1996, \aaps, 120, 157

\bibitem[{{Gray} \& {Corbally}(2009)}]{Gray2009}
{Gray}, R.~O. \& {Corbally}, J., C. 2009, {Stellar Spectral Classification}
  (Princeton University Press)

\bibitem[{{Haguenauer} {et~al.}(2010){Haguenauer}, {Alonso}, {Bourget},
  {Brillant}, {Gitton}, {Guisard}, {Poupar}, {Schuhler}, {Abuter}, {Andolfato},
  {Blanchard}, {Berger}, {Cortes}, {D{\'e}rie}, {Delplancke}, {di Lieto},
  {Dupuy}, {Gilli}, {Glindemann}, {Guniat}, {Huedepohl}, {Kaufer}, {Le
  Bouquin}, {L{\'e}v{\^e}que}, {M{\'e}nardi}, {M{\'e}rand}, {Morel},
  {Percheron}, {Phan Duc}, {Pino}, {Ramirez}, {Rengaswamy}, {Richichi},
  {Rivinius}, {Sahlmann}, {Schoeller}, {Schmid}, {Stefl}, {Valdes}, {van
  Belle}, {Wehner}, \& {Wittkowski}}]{Haguenauer2010}
{Haguenauer}, P., {Alonso}, J., {Bourget}, P., {et~al.} 2010, in Society of
  Photo-Optical Instrumentation Engineers (SPIE) Conference Series, Vol. 7734,
  Society of Photo-Optical Instrumentation Engineers (SPIE) Conference Series

\bibitem[{{Halbedel}(1991)}]{Halbedel1991}
{Halbedel}, E.~M. 1991, Information Bulletin on Variable Stars, 3585, 1

\bibitem[{{Harrington} \& {Kuhn}(2007)}]{Harrington2007}
{Harrington}, D.~M. \& {Kuhn}, J.~R. 2007, \apjl, 667, L89

\bibitem[{{Harrington} \& {Kuhn}(2009)}]{Harrington2009}
{Harrington}, D.~M. \& {Kuhn}, J.~R. 2009, \apjs, 180, 138

\bibitem[{{Heber}(2009)}]{Heber2009}
{Heber}, U. 2009, \araa, 47, 211

\bibitem[{{Horne} \& {Marsh}(1986)}]{Horne1986}
{Horne}, K. \& {Marsh}, T.~R. 1986, \mnras, 218, 761

\bibitem[{{Hosokawa} {et~al.}(2010){Hosokawa}, {Yorke}, \&
  {Omukai}}]{Hosokawa2010}
{Hosokawa}, T., {Yorke}, H.~W., \& {Omukai}, K. 2010, \apj, 721, 478

\bibitem[{{Hutsem\'{e}kers}(1985)}]{Hutsemekers1985}
{Hutsem\'{e}kers}, D. 1985, \aaps, 60, 373

\bibitem[{{Ilee} {et~al.}(2013){Ilee}, {Wheelwright}, {Oudmaijer}, {de Wit},
  {Maud}, {Hoare}, {Lumsden}, {Moore}, {Urquhart}, \& {Mottram}}]{Ilee2013}
{Ilee}, J.~D., {Wheelwright}, H.~E., {Oudmaijer}, R.~D., {et~al.} 2013, \mnras,
  429, 2960

\bibitem[{{Jaschek} \& {Andrillat}(1998)}]{Jaschek1998}
{Jaschek}, C. \& {Andrillat}, Y. 1998, \aaps, 128, 475

\bibitem[{{Jaschek} {et~al.}(1993){Jaschek}, {Jaschek}, \&
  {Andrillat}}]{Jaschek1993}
{Jaschek}, M., {Jaschek}, C., \& {Andrillat}, Y. 1993, \aaps, 97, 781

\bibitem[{{K{\'o}sp{\'a}l} {et~al.}(2013){K{\'o}sp{\'a}l}, {{\'A}brah{\'a}m},
  {Acosta-Pulido}, {Ar{\'e}valo Morales}, {Balog}, {Carnerero}, {Szegedi-Elek},
  {Farkas}, {Henning}, {Kelemen}, {Kov{\'a}cs}, {Kun}, {Marton},
  {M{\'e}sz{\'a}ros}, {Mo{\'o}r}, {P{\'a}l}, {S{\'a}rneczky}, {Szak{\'a}ts},
  {Szalai}, {Szing}, {T{\'o}th}, {Turner}, \& {Vida}}]{Kospal2013}
{K{\'o}sp{\'a}l}, {\'A}., {{\'A}brah{\'a}m}, P., {Acosta-Pulido}, J.~A.,
  {et~al.} 2013, ArXiv e-prints

\bibitem[{{Kraus} {et~al.}(2013){Kraus}, {Oksala}, {Nickeler}, {Muratore},
  {Borges Fernandes}, {Aret}, {Cidale}, \& {de Wit}}]{Kraus2013}
{Kraus}, M., {Oksala}, M.~E., {Nickeler}, D.~H., {et~al.} 2013, \aap, 549, A28

\bibitem[{{Kraus} {et~al.}(2012){Kraus}, {Calvet}, {Hartmann}, {Hofmann},
  {Kreplin}, {Monnier}, \& {Weigelt}}]{Kraus2012}
{Kraus}, S., {Calvet}, N., {Hartmann}, L., {et~al.} 2012, \apj, 752, 11

\bibitem[{{Kraus} {et~al.}(2008){Kraus}, {Hofmann}, {Benisty}, {Berger},
  {Chesneau}, {Isella}, {Malbet}, {Meilland}, {Nardetto}, {Natta}, {Preibisch},
  {Schertl}, {Smith}, {Stee}, {Tatulli}, {Testi}, \& {Weigelt}}]{Kraus2008b}
{Kraus}, S., {Hofmann}, K.-H., {Benisty}, M., {et~al.} 2008, \aap, 489, 1157

\bibitem[{{Krumholz} {et~al.}(2009){Krumholz}, {Klein}, {McKee}, {Offner}, \&
  {Cunningham}}]{Krumholz2009}
{Krumholz}, M.~R., {Klein}, R.~I., {McKee}, C.~F., {Offner}, S.~S.~R., \&
  {Cunningham}, A.~J. 2009, Science, 323, 754

\bibitem[{{Kurucz}(1993)}]{Kurucz1993}
{Kurucz}, R.~L. 1993, VizieR Online Data Catalog, 6039, 0

\bibitem[{{Lachaume}(2003)}]{Lachaume2003}
{Lachaume}, R. 2003, \aap, 400, 795

\bibitem[{{Lagage} {et~al.}(2006){Lagage}, {Doucet}, {Pantin}, {Habart},
  {Duch{\^e}ne}, {M{\'e}nard}, {Pinte}, {Charnoz}, \& {Pel}}]{Lagage2006}
{Lagage}, P.-O., {Doucet}, C., {Pantin}, E., {et~al.} 2006, Science, 314, 621

\bibitem[{{Lamers} \& {Pauldrach}(1991)}]{Lamers1991}
{Lamers}, H.~J.~G. \& {Pauldrach}, A.~W.~A. 1991, \aap, 244, L5

\bibitem[{{Lamers} {et~al.}(1998){Lamers}, {Zickgraf}, {de Winter}, {Houziaux},
  \& {Zorec}}]{Lamers1998}
{Lamers}, H.~J.~G.~L.~M., {Zickgraf}, F.-J., {de Winter}, D., {Houziaux}, L.,
  \& {Zorec}, J. 1998, \aap, 340, 117

\bibitem[{{Le Bouquin} {et~al.}(2008){Le Bouquin}, {Bauvir}, {Haguenauer},
  {Sch{\"o}ller}, {Rantakyr{\"o}}, \& {Menardi}}]{lebouquin08}
{Le Bouquin}, J.-B., {Bauvir}, B., {Haguenauer}, P., {et~al.} 2008, \aap, 481,
  553

\bibitem[{{Maddalena} {et~al.}(1986){Maddalena}, {Morris}, {Moscowitz}, \&
  {Thaddeus}}]{Maddalena1986}
{Maddalena}, R.~J., {Morris}, M., {Moscowitz}, J., \& {Thaddeus}, P. 1986,
  \apj, 303, 375

\bibitem[{{Maheswar} {et~al.}(2002){Maheswar}, {Manoj}, \&
  {Bhatt}}]{Maheswar2002}
{Maheswar}, G., {Manoj}, P., \& {Bhatt}, H.~C. 2002, \aap, 387, 1003

\bibitem[{{Malbet} {et~al.}(2007){Malbet}, {Benisty}, {de Wit}, {Kraus},
  {Meilland}, {Millour}, {Tatulli}, {Berger}, {Chesneau}, {Hofmann}, {Isella},
  {Natta}, {Petrov}, {Preibisch}, {Stee}, {Testi}, {Weigelt}, {Antonelli},
  {Beckmann}, {Bresson}, {Chelli}, {Dugu{\'e}}, {Duvert}, {Gennari},
  {Gl{\"u}ck}, {Kern}, {Lagarde}, {Le Coarer}, {Lisi}, {Perraut}, {Puget},
  {Rantakyr{\"o}}, {Robbe-Dubois}, {Roussel}, {Zins}, {Accardo}, {Acke},
  {Agabi}, {Altariba}, {Arezki}, {Aristidi}, {Baffa}, {Behrend}, {Bl{\"o}cker},
  {Bonhomme}, {Busoni}, {Cassaing}, {Clausse}, {Colin}, {Connot},
  {Delboulb{\'e}}, {Domiciano de Souza}, {Driebe}, {Feautrier}, {Ferruzzi},
  {Forveille}, {Fossat}, {Foy}, {Fraix-Burnet}, {Gallardo}, {Giani}, {Gil},
  {Glentzlin}, {Heiden}, {Heininger}, {Hernandez Utrera}, {Kamm}, {Kiekebusch},
  {Le Contel}, {Le Contel}, {Lesourd}, {Lopez}, {Lopez}, {Magnard}, {Marconi},
  {Mars}, {Martinot-Lagarde}, {Mathias}, {M{\`e}ge}, {Monin}, {Mouillet},
  {Mourard}, {Nussbaum}, {Ohnaka}, {Pacheco}, {Perrier}, {Rabbia}, {Rebattu},
  {Reynaud}, {Richichi}, {Robini}, {Sacchettini}, {Schertl}, {Sch{\"o}ller},
  {Solscheid}, {Spang}, {Stefanini}, {Tallon}, {Tallon-Bosc}, {Tasso},
  {Vakili}, {von der L{\"u}he}, {Valtier}, {Vannier}, \&
  {Ventura}}]{Malbet2007}
{Malbet}, F., {Benisty}, M., {de Wit}, W.-J., {et~al.} 2007, \aap, 464, 43

\bibitem[{{Meilland} {et~al.}(2012){Meilland}, {Millour}, {Kanaan}, {Stee},
  {Petrov}, {Hofmann}, {Natta}, \& {Perraut}}]{Meilland2012}
{Meilland}, A., {Millour}, F., {Kanaan}, S., {et~al.} 2012, \aap, 538, A110

\bibitem[{{M{\'e}rand} {et~al.}(2005){M{\'e}rand}, {Bord{\'e}}, \& {Coud{\'e}
  du Foresto}}]{Merand2005}
{M{\'e}rand}, A., {Bord{\'e}}, P., \& {Coud{\'e} du Foresto}, V. 2005, \aap,
  433, 1155

\bibitem[{{Merrill}(1931)}]{Merrill1931}
{Merrill}, P.~W. 1931, \apj, 73, 348

\bibitem[{{Merrill} \& {Burwell}(1933)}]{Merrill1933}
{Merrill}, P.~W. \& {Burwell}, C.~G. 1933, \apj, 78, 87

\bibitem[{{Millour} {et~al.}(2009){Millour}, {Chesneau}, {Borges Fernandes},
  {Meilland}, {Mars}, {Benoist}, {Thi{\'e}baut}, {Stee}, {Hofmann}, {Baron},
  {Young}, {Bendjoya}, {Carciofi}, {Domiciano de Souza}, {Driebe}, {Jankov},
  {Kervella}, {Petrov}, {Robbe-Dubois}, {Vakili}, {Waters}, \&
  {Weigelt}}]{Millour2009}
{Millour}, F., {Chesneau}, O., {Borges Fernandes}, M., {et~al.} 2009, \aap,
  507, 317

\bibitem[{{Miroshnichenko}(2007)}]{Miroshnichenko2007}
{Miroshnichenko}, A.~S. 2007, \apj, 667, 497

\bibitem[{{Modigliani} {et~al.}(2010){Modigliani}, {Goldoni}, {Royer},
  {Haigron}, {Guglielmi}, {Fran{\c c}ois}, {Horrobin}, {Bristow}, {Vernet},
  {Moehler}, {Kerber}, {Ballester}, {Mason}, \& {Christensen}}]{Modigliani2010}
{Modigliani}, A., {Goldoni}, P., {Royer}, F., {et~al.} 2010, in Society of
  Photo-Optical Instrumentation Engineers (SPIE) Conference Series, Vol. 7737

\bibitem[{{Morrison} \& {Beaver}(1995)}]{Morrison1995}
{Morrison}, N.~D. \& {Beaver}, M. 1995, in Bulletin of the American
  Astronomical Society, Vol.~27, American Astronomical Society Meeting
  Abstracts \#186, 825

\bibitem[{{Mourard} {et~al.}(2009){Mourard}, {Clausse}, {Marcotto}, {{Perraut},
  K.}, {Tallon-Bosc}, {B{\'e}rio}, {Blazit}, {Bonneau}, {Bosio}, {Bresson},
  {Chesneau}, {Delaa}, {H{\'e}nault}, {Hughes}, {Lagarde}, {Merlin}, {Roussel},
  {Spang}, {Stee}, {Tallon}, {Antonelli}, {Foy}, {Kervella}, {Petrov},
  {Thiebaut}, {Vakili}, {McAlister}, {ten Brummelaar}, {Sturmann}, {Sturmann},
  {Turner}, {Farrington}, \& {Goldfinger}}]{Mourard2009}
{Mourard}, D., {Clausse}, J.~M., {Marcotto}, A., {et~al.} 2009, Astronomy \&
  Astrophysics, 508, 1073

\bibitem[{{Ochsendorf} {et~al.}(2011){Ochsendorf}, {Ellerbroek}, {Chini},
  {Hartoog}, {Hoffmeister}, {Waters}, \& {Kaper}}]{Ochsendorf2011}
{Ochsendorf}, B.~B., {Ellerbroek}, L.~E., {Chini}, R., {et~al.} 2011, \aap,
  536, L1

\bibitem[{{Oudmaijer} \& {Drew}(1999)}]{Oudmaijer1999}
{Oudmaijer}, R.~D. \& {Drew}, J.~E. 1999, \mnras, 305, 166

\bibitem[{{Oudmaijer} {et~al.}(2011){Oudmaijer}, {Wheelwright}, {Carciofi},
  {Bjorkman}, \& {Bjorkman}}]{Oudmaijer2011}
{Oudmaijer}, R.~D., {Wheelwright}, H.~E., {Carciofi}, A.~C., {Bjorkman}, J.~E.,
  \& {Bjorkman}, K.~S. 2011, in IAU Symposium, Vol. 272, IAU Symposium, ed.
  C.~{Neiner}, G.~{Wade}, G.~{Meynet}, \& G.~{Peters}, 418--419

\bibitem[{{Pauls} {et~al.}(2005){Pauls}, {Young}, {Cotton}, \&
  {Monnier}}]{pauls05}
{Pauls}, T.~A., {Young}, J.~S., {Cotton}, W.~D., \& {Monnier}, J.~D. 2005,
  \pasp, 117, 1255

\bibitem[{{Perryman} {et~al.}(2001){Perryman}, {de Boer}, {Gilmore}, {H{\o}g},
  {Lattanzi}, {Lindegren}, {Luri}, {Mignard}, {Pace}, \& {de
  Zeeuw}}]{Perryman2001}
{Perryman}, M.~A.~C., {de Boer}, K.~S., {Gilmore}, G., {et~al.} 2001, \aap,
  369, 339

\bibitem[{{Petrov} {et~al.}(2007){Petrov}, {Malbet}, {Weigelt}, {Antonelli},
  {Beckmann}, {Bresson}, {Chelli}, {Dugu{\'e}}, {Duvert}, {Gennari},
  {Gl{\"u}ck}, {Kern}, {Lagarde}, {Le Coarer}, {Lisi}, {Millour}, {Perraut},
  {Puget}, {Rantakyr{\"o}}, {Robbe-Dubois}, {Roussel}, {Salinari}, {Tatulli},
  {Zins}, {Accardo}, {Acke}, {Agabi}, {Altariba}, {Arezki}, {Aristidi},
  {Baffa}, {Behrend}, {Bl{\"o}cker}, {Bonhomme}, {Busoni}, {Cassaing},
  {Clausse}, {Colin}, {Connot}, {Delboulb{\'e}}, {Domiciano de Souza},
  {Driebe}, {Feautrier}, {Ferruzzi}, {Forveille}, {Fossat}, {Foy},
  {Fraix-Burnet}, {Gallardo}, {Giani}, {Gil}, {Glentzlin}, {Heiden},
  {Heininger}, {Hernandez Utrera}, {Hofmann}, {Kamm}, {Kiekebusch}, {Kraus},
  {Le Contel}, {Le Contel}, {Lesourd}, {Lopez}, {Lopez}, {Magnard}, {Marconi},
  {Mars}, {Martinot-Lagarde}, {Mathias}, {M{\`e}ge}, {Monin}, {Mouillet},
  {Mourard}, {Nussbaum}, {Ohnaka}, {Pacheco}, {Perrier}, {Rabbia}, {Rebattu},
  {Reynaud}, {Richichi}, {Robini}, {Sacchettini}, {Schertl}, {Sch{\"o}ller},
  {Solscheid}, {Spang}, {Stee}, {Stefanini}, {Tallon}, {Tallon-Bosc}, {Tasso},
  {Testi}, {Vakili}, {von der L{\"u}he}, {Valtier}, {Vannier}, \&
  {Ventura}}]{petrov07}
{Petrov}, R.~G., {Malbet}, F., {Weigelt}, G., {et~al.} 2007, \aap, 464, 1

\bibitem[{{Pogodin}(1997)}]{Pogodin1997}
{Pogodin}, M.~A. 1997, \aap, 317, 185

\bibitem[{{Pontoppidan} {et~al.}(2011){Pontoppidan}, {Blake}, \&
  {Smette}}]{Pontoppidan2011}
{Pontoppidan}, K.~M., {Blake}, G.~A., \& {Smette}, A. 2011, \apj, 733, 84

\bibitem[{{Puls} {et~al.}(2005){Puls}, {Urbaneja}, {Venero}, {Repolust},
  {Springmann}, {Jokuthy}, \& {Mokiem}}]{Puls2005}
{Puls}, J., {Urbaneja}, M.~A., {Venero}, R., {et~al.} 2005, \aap, 435, 669

\bibitem[{{Schaller} {et~al.}(1992){Schaller}, {Schaerer}, {Meynet}, \&
  {Maeder}}]{Schaller1992}
{Schaller}, G., {Schaerer}, D., {Meynet}, G., \& {Maeder}, A. 1992, \aaps, 96,
  269

\bibitem[{{Sch{\"o}ller}(2007)}]{scholler07}
{Sch{\"o}ller}, M. 2007, New Astronomy Review, 51, 628

\bibitem[{{Sitko} {et~al.}(2008){Sitko}, {Carpenter}, {Kimes}, {Wilde},
  {Lynch}, {Russell}, {Rudy}, {Mazuk}, {Venturini}, {Puetter}, {Grady},
  {Polomski}, {Wisnewski}, {Brafford}, {Hammel}, \& {Perry}}]{Sitko2008}
{Sitko}, M.~L., {Carpenter}, W.~J., {Kimes}, R.~L., {et~al.} 2008, \apj, 678,
  1070

\bibitem[{{Sitko} {et~al.}(2004){Sitko}, {Carpenter}, {Lynch}, {Russell},
  {Grady}, {Brafford}, \& {Wooden}}]{Sitko2004}
{Sitko}, M.~L., {Carpenter}, W.~J., {Lynch}, D.~K., {et~al.} 2004, in Bulletin
  of the American Astronomical Society, Vol.~36, American Astronomical Society
  Meeting Abstracts, 1363

\bibitem[{{Slettebak}(1976)}]{Slettebak1976}
{Slettebak}, A., ed. 1976, IAU Symposium, Vol.~70, {Be and shell stars;
  Proceedings of the Merrill-McLaughlin Memorial Symposium, BASS River, Mass.,
  September 15-18, 1975}

\bibitem[{{Sturmann} {et~al.}(2010){Sturmann}, {Ten Brummelaar}, {Sturmann}, \&
  {McAlister}}]{Sturmann2010}
{Sturmann}, J., {Ten Brummelaar}, T., {Sturmann}, L., \& {McAlister}, H.~A.
  2010, in Society of Photo-Optical Instrumentation Engineers (SPIE) Conference
  Series, Vol. 7734, Society of Photo-Optical Instrumentation Engineers (SPIE)
  Conference Series

\bibitem[{{Tatulli} {et~al.}(2007{\natexlab{a}}){Tatulli}, {Isella}, {Natta},
  {Testi}, {Marconi}, {Malbet}, {Stee}, {Petrov}, {Millour}, {Chelli},
  {Duvert}, {Antonelli}, {Beckmann}, {Bresson}, {Dugu{\'e}}, {Gennari},
  {Gl{\"u}ck}, {Kern}, {Lagarde}, {Le Coarer}, {Lisi}, {Perraut}, {Puget},
  {Rantakyr{\"o}}, {Robbe-Dubois}, {Roussel}, {Weigelt}, {Zins}, {Accardo},
  {Acke}, {Agabi}, {Altariba}, {Arezki}, {Aristidi}, {Baffa}, {Behrend},
  {Bl{\"o}cker}, {Bonhomme}, {Busoni}, {Cassaing}, {Clausse}, {Colin},
  {Connot}, {Delboulb{\'e}}, {Domiciano de Souza}, {Driebe}, {Feautrier},
  {Ferruzzi}, {Forveille}, {Fossat}, {Foy}, {Fraix-Burnet}, {Gallardo},
  {Giani}, {Gil}, {Glentzlin}, {Heiden}, {Heininger}, {Hernandez Utrera},
  {Hofmann}, {Kamm}, {Kiekebusch}, {Kraus}, {Le Contel}, {Le Contel},
  {Lesourd}, {Lopez}, {Lopez}, {Magnard}, {Mars}, {Martinot-Lagarde},
  {Mathias}, {M{\`e}ge}, {Monin}, {Mouillet}, {Mourard}, {Nussbaum}, {Ohnaka},
  {Pacheco}, {Perrier}, {Rabbia}, {Rebattu}, {Reynaud}, {Richichi}, {Robini},
  {Sacchettini}, {Schertl}, {Sch{\"o}ller}, {Solscheid}, {Spang}, {Stefanini},
  {Tallon}, {Tallon-Bosc}, {Tasso}, {Vakili}, {von der L{\"u}he}, {Valtier},
  {Vannier}, \& {Ventura}}]{Tatulli2007b}
{Tatulli}, E., {Isella}, A., {Natta}, A., {et~al.} 2007{\natexlab{a}}, \aap,
  464, 55

\bibitem[{{Tatulli} {et~al.}(2007{\natexlab{b}}){Tatulli}, {Millour}, {Chelli},
  {Duvert}, {Acke}, {Hernandez Utrera}, {Hofmann}, {Kraus}, {Malbet},
  {M{\`e}ge}, {Petrov}, {Vannier}, {Zins}, {Antonelli}, {Beckmann}, {Bresson},
  {Dugu{\'e}}, {Gennari}, {Gl{\"u}ck}, {Kern}, {Lagarde}, {Le Coarer}, {Lisi},
  {Perraut}, {Puget}, {Rantakyr{\"o}}, {Robbe-Dubois}, {Roussel}, {Weigelt},
  {Accardo}, {Agabi}, {Altariba}, {Arezki}, {Aristidi}, {Baffa}, {Behrend},
  {Bl{\"o}cker}, {Bonhomme}, {Busoni}, {Cassaing}, {Clausse}, {Colin},
  {Connot}, {Delboulb{\'e}}, {Domiciano de Souza}, {Driebe}, {Feautrier},
  {Ferruzzi}, {Forveille}, {Fossat}, {Foy}, {Fraix-Burnet}, {Gallardo},
  {Giani}, {Gil}, {Glentzlin}, {Heiden}, {Heininger}, {Kamm}, {Kiekebusch}, {Le
  Contel}, {Le Contel}, {Lesourd}, {Lopez}, {Lopez}, {Magnard}, {Marconi},
  {Mars}, {Martinot-Lagarde}, {Mathias}, {Monin}, {Mouillet}, {Mourard},
  {Nussbaum}, {Ohnaka}, {Pacheco}, {Perrier}, {Rabbia}, {Rebattu}, {Reynaud},
  {Richichi}, {Robini}, {Sacchettini}, {Schertl}, {Sch{\"o}ller}, {Solscheid},
  {Spang}, {Stee}, {Stefanini}, {Tallon}, {Tallon-Bosc}, {Tasso}, {Testi},
  {Vakili}, {von der L{\"u}he}, {Valtier}, \& {Ventura}}]{Tatulli2007a}
{Tatulli}, E., {Millour}, F., {Chelli}, A., {et~al.} 2007{\natexlab{b}}, \aap,
  464, 29

\bibitem[{{ten Brummelaar} {et~al.}(2005){ten Brummelaar}, {McAlister},
  {Ridgway}, {Bagnuolo}, {Turner}, {Sturmann}, {Sturmann}, {Berger}, {Ogden},
  {Cadman}, {Hartkopf}, {Hopper}, \& {Shure}}]{TenBrummelaar2005}
{ten Brummelaar}, T.~A., {McAlister}, H.~A., {Ridgway}, S.~T., {et~al.} 2005,
  \apj, 628, 453

\bibitem[{{van Leeuwen}(2007)}]{VanLeeuwen2007}
{van Leeuwen}, F. 2007, \aap, 474, 653

\bibitem[{{Vernet} {et~al.}(2011){Vernet}, {Dekker}, {D'Odorico}, {Kaper},
  {Kjaergaard}, {Hammer}, {Randich}, {Zerbi}, {Groot}, {Hjorth}, {Guinouard},
  {Navarro}, {Adolfse}, {Albers}, {Amans}, {Andersen}, {Andersen}, {Binetruy},
  {Bristow}, {Castillo}, {Chemla}, {Christensen}, {Conconi}, {Conzelmann},
  {Dam}, {De Caprio}, {De Ugarte Postigo}, {Delabre}, {Di Marcantonio},
  {Downing}, {Elswijk}, {Finger}, {Fischer}, {Flores}, {Francois}, {Goldoni},
  {Guglielmi}, {Haigron}, {Hanenburg}, {Hendriks}, {Horrobin}, {Horville},
  {Jessen}, {Kerber}, {Kern}, {Kiekebusch}, {Kleszcz}, {Klougart}, {Kragt},
  {Larsen}, {Lizon}, {Lucuix}, {Mainieri}, {Manuputy}, {Martayan}, {Mason},
  {Mazzoleni}, {Michaelsen}, {Modigliani}, {Moehler}, {M{\o}ller}, {Norup
  S{\o}rensen}, {N{\o}rregaard}, {Peroux}, {Patat}, {Pena}, {Pragt}, {Reinero},
  {Riga}, {Riva}, {Roelfsema}, {Royer}, {Sacco}, {Santin}, {Schoenmaker},
  {Spano}, {Sweers}, {Ter Horst}, {Tintori}, {Tromp}, {van Dael}, {van der
  Vliet}, {Venema}, {Vidali}, {Vinther}, {Vola}, {Winters}, {Wistisen},
  {Wulterkens}, \& {Zacchei}}]{Vernet2011}
{Vernet}, J., {Dekker}, H., {D'Odorico}, S., {et~al.} 2011, ArXiv e-prints

\bibitem[{{Wang} {et~al.}(2012){Wang}, {Weigelt}, {Kreplin}, {Hofmann},
  {Kraus}, {Miroshnichenko}, {Schertl}, {Chelli}, {Domiciano de Souza},
  {Massi}, \& {Robbe-Dubois}}]{Wang2012}
{Wang}, Y., {Weigelt}, G., {Kreplin}, A., {et~al.} 2012, \aap, 545, L10

\bibitem[{{Weigelt} {et~al.}(2011){Weigelt}, {Grinin}, {Groh}, {Hofmann},
  {Kraus}, {Miroshnichenko}, {Schertl}, {Tambovtseva}, {Benisty}, {Driebe},
  {Lagarde}, {Malbet}, {Meilland}, {Petrov}, \& {Tatulli}}]{Weigelt2011}
{Weigelt}, G., {Grinin}, V.~P., {Groh}, J.~H., {et~al.} 2011, \aap, 527, A103

\bibitem[{{Weigelt} {et~al.}(2007){Weigelt}, {Kraus}, {Driebe}, {Petrov},
  {Hofmann}, {Millour}, {Chesneau}, {Schertl}, {Malbet}, {Hillier}, {Gull},
  {Davidson}, {Domiciano de Souza}, {Antonelli}, {Beckmann}, {Bresson},
  {Chelli}, {Dugu{\'e}}, {Duvert}, {Gennari}, {Gl{\"u}ck}, {Kern}, {Lagarde},
  {Le Coarer}, {Lisi}, {Perraut}, {Puget}, {Rantakyr{\"o}}, {Robbe-Dubois},
  {Roussel}, {Tatulli}, {Zins}, {Accardo}, {Acke}, {Agabi}, {Altariba},
  {Arezki}, {Aristidi}, {Baffa}, {Behrend}, {Bl{\"o}cker}, {Bonhomme},
  {Busoni}, {Cassaing}, {Clausse}, {Colin}, {Connot}, {Delboulb{\'e}},
  {Feautrier}, {Ferruzzi}, {Forveille}, {Fossat}, {Foy}, {Fraix-Burnet},
  {Gallardo}, {Giani}, {Gil}, {Glentzlin}, {Heiden}, {Heininger}, {Hernandez
  Utrera}, {Kamm}, {Kiekebusch}, {Le Contel}, {Le Contel}, {Lesourd}, {Lopez},
  {Lopez}, {Magnard}, {Marconi}, {Mars}, {Martinot-Lagarde}, {Mathias},
  {M{\`e}ge}, {Monin}, {Mouillet}, {Mourard}, {Nussbaum}, {Ohnaka}, {Pacheco},
  {Perrier}, {Rabbia}, {Rebattu}, {Reynaud}, {Richichi}, {Robini},
  {Sacchettini}, {Sch{\"o}ller}, {Solscheid}, {Spang}, {Stee}, {Stefanini},
  {Tallon}, {Tallon-Bosc}, {Tasso}, {Testi}, {Vakili}, {von der L{\"u}he},
  {Valtier}, {Vannier}, {Ventura}, {Weis}, \& {Wittkowski}}]{Weigelt2007}
{Weigelt}, G., {Kraus}, S., {Driebe}, T., {et~al.} 2007, \aap, 464, 87

\bibitem[{{Wheelwright} {et~al.}(2012{\natexlab{a}}){Wheelwright}, {Bjorkman},
  {Oudmaijer}, {Carciofi}, {Bjorkman}, \& {Porter}}]{Wheelwright2012b}
{Wheelwright}, H.~E., {Bjorkman}, J.~E., {Oudmaijer}, R.~D., {et~al.}
  2012{\natexlab{a}}, \mnras, 423, L11

\bibitem[{{Wheelwright} {et~al.}(2012{\natexlab{b}}){Wheelwright}, {de Wit},
  {Oudmaijer}, \& {Vink}}]{Wheelwright2012a}
{Wheelwright}, H.~E., {de Wit}, W.~J., {Oudmaijer}, R.~D., \& {Vink}, J.~S.
  2012{\natexlab{b}}, \aap, 538, A6

\bibitem[{{Wheelwright} {et~al.}(2012{\natexlab{c}}){Wheelwright}, {de Wit},
  {Weigelt}, {Oudmaijer}, \& {Ilee}}]{Wheelwright2012c}
{Wheelwright}, H.~E., {de Wit}, W.~J., {Weigelt}, G., {Oudmaijer}, R.~D., \&
  {Ilee}, J.~D. 2012{\natexlab{c}}, \aap, 543, A77

\bibitem[{{Wheelwright} {et~al.}(2011){Wheelwright}, {Vink}, {Oudmaijer}, \&
  {Drew}}]{Wheelwright2011}
{Wheelwright}, H.~E., {Vink}, J.~S., {Oudmaijer}, R.~D., \& {Drew}, J.~E. 2011,
  \aap, 532, A28

\bibitem[{{Wheelwright} {et~al.}(2013){Wheelwright}, {Weigelt}, {Caratti o
  Garatti}, \& {Garcia Lopez}}]{Wheelwright2013}
{Wheelwright}, H.~E., {Weigelt}, G., {Caratti o Garatti}, A., \& {Garcia
  Lopez}, R. 2013, \aap, 558, A116

\bibitem[{{Yudin} \& {Evans}(1998)}]{Yudin1998}
{Yudin}, R.~V. \& {Evans}, A. 1998, \aaps, 131, 401

\bibitem[{{Zickgraf}(1998)}]{Zickgraf1998}
{Zickgraf}, F.-J. 1998, in Astrophysics and Space Science Library, Vol. 233,
  B[e] stars, ed. A.~M. {Hubert} \& C.~{Jaschek}, 1

\end{thebibliography}

\clearpage

\appendix 



\section{Supplementary figures}

\label{sec:sed}
\noindent\begin{minipage}{\textwidth}
   \centering
   \includegraphics[width=0.5\textwidth]{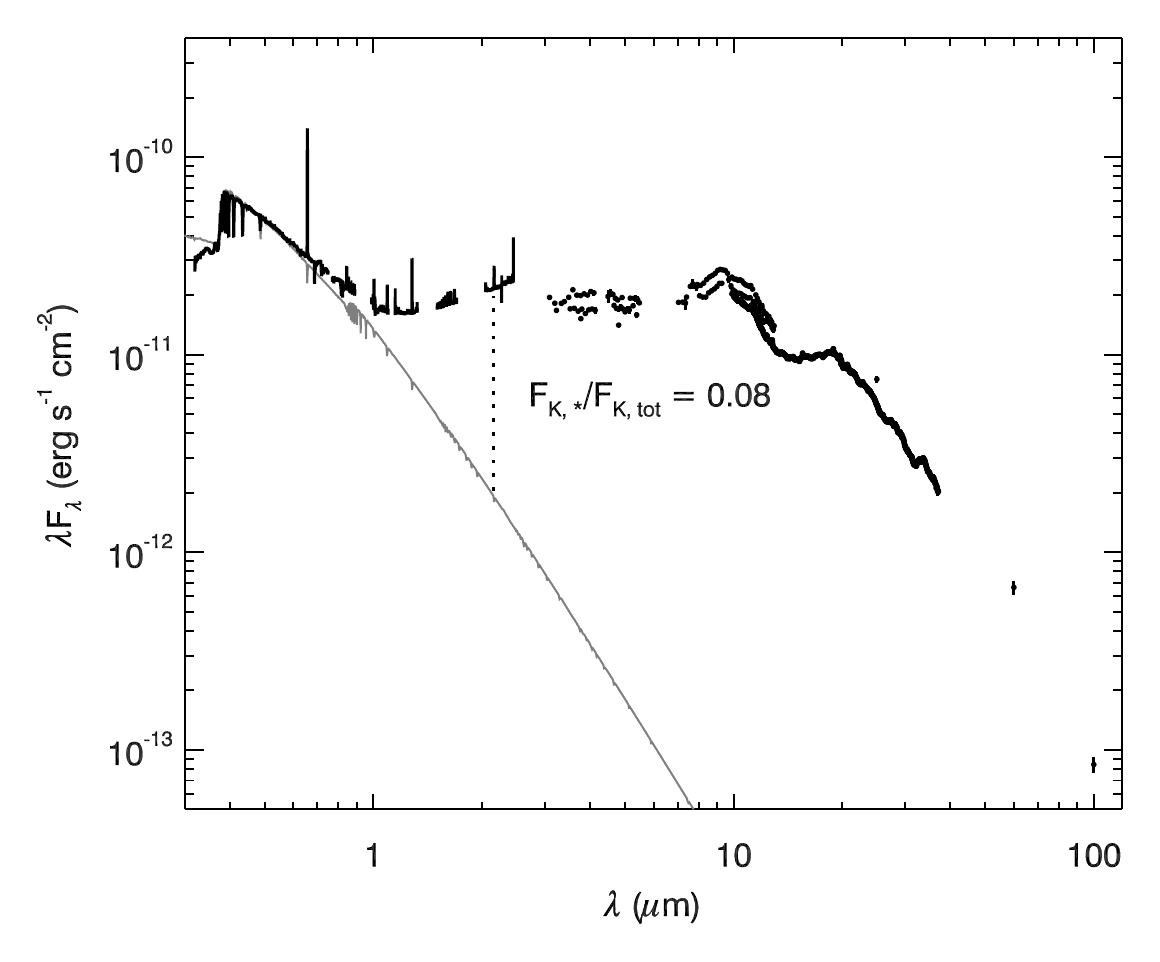}
   \captionof{figure}{Spectral energy distribution; data from X-shooter (black line), other data from \citet[][black symbols]{Sitko2004}; model photosphere at 13000 K \citep[][gray line]{Kurucz1993} and at $d=500$~pc. The model is reddened with $A_V=0.4$~mag using the extinction law by \citet{Cardelli1989}. The vertical dotted line corresponds to the stellar-to-total flux ratio in the $K$-band ($\lambda=2.16~\mu$m), $f=0.08$.}
   \label{fig:sed}
\vspace{1cm}
 \includegraphics[width=0.99\textwidth, page=2]{astrometry.pdf}
\captionof{figure}{Astrometric solution, $\vec{P}(\lambda)$, (yellow line) overplotted on the differential phase observations which were not included in the fit. The error bars correspond to the 1$\sigma$ spread in the continuum region.}
\label{fig:astromtetry:misc}
\end{minipage}
\clearpage

\begin{figure*}
   \centering
   \includegraphics[height=0.85\textwidth, trim=0cm 13.5cm 0cm 0cm, clip=true, angle=90, origin=c]{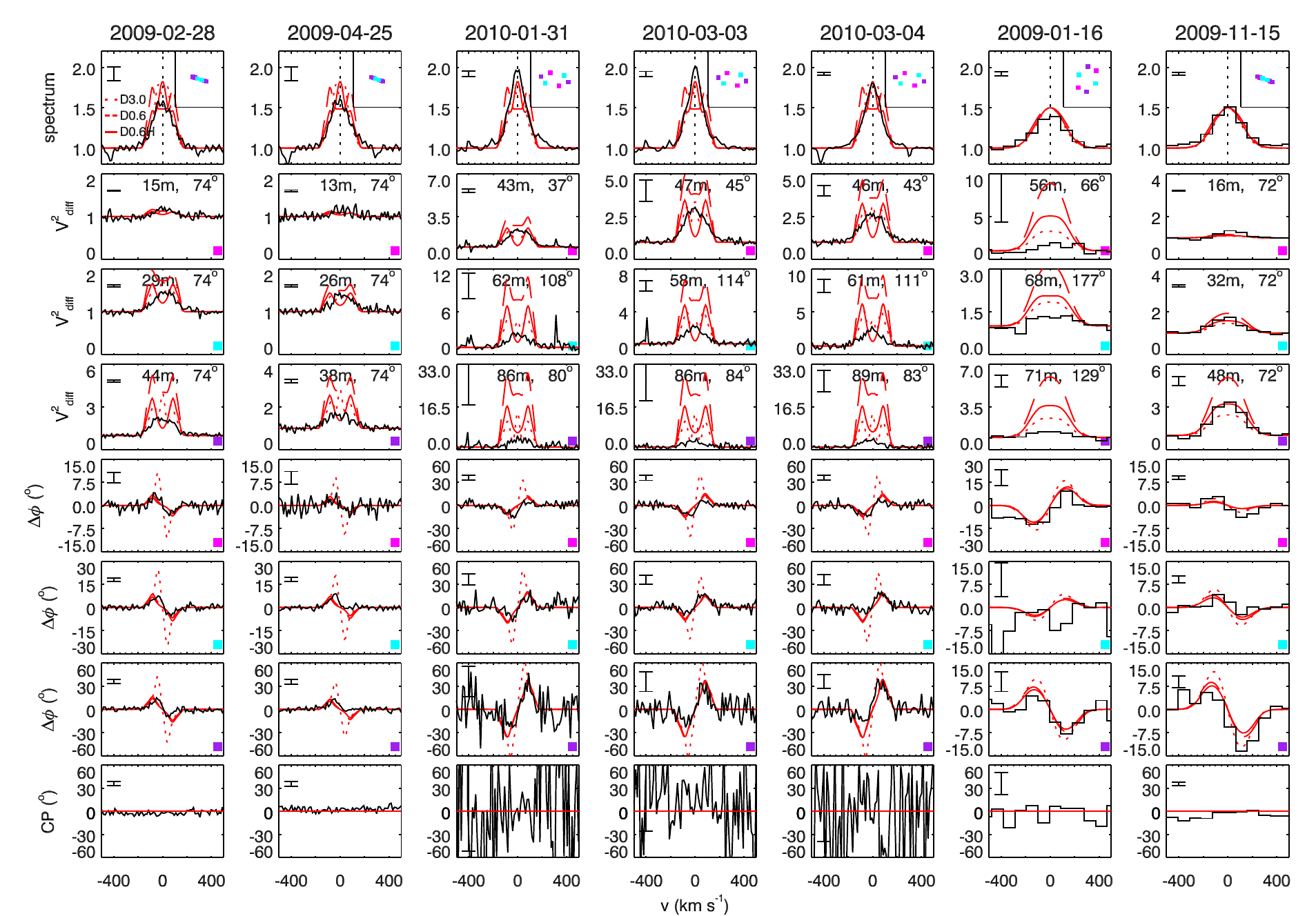}
      \caption{Same as Fig.~\ref{fig:data}, but with $V^2_{\rm diff}$ plotted instead of the absolute visibility. Overplotted are the observables calculated from models D0.6 ($R_{\rm out}=0.6$~au, red line), D3.0 ($R_{\rm out}=3$~au, blue line), and D0.6H ($R_{\rm out}=0.6$~au and spherical halo, green line).} 
   \label{fig:data_models_select}
\end{figure*}
\clearpage

\addtocounter{figure}{-1}

\begin{figure*}
   \centering
   \includegraphics[height=0.85\textwidth, trim=0cm 0cm 0cm 13.5cm, clip=true, angle=270, origin=c]{fig_data_models_select_best3.pdf}
      \caption{Continued.} 
\end{figure*}

\end{document}